# A general systems theory for rain formation in warm clouds

A.M.Selvam[1]

*B1 Aradhana, 42/2A Shivajinagar, Pune 411005, India*

*Email: amselvam@gmail.com*
*Websites: http://amselvam.webs.com*
*http://amselvam.tripod.com/index.html*

---

[1] Retired Deputy Director, Indian Institute of Tropical Meteorology, Pune 411008, India
Corresponding author permanent address: Dr.Mrs.A.M.Selvam,
B1 Aradhana, 42/2A Shivajinagar, Pune 411005, India. Tel.(Res.) 091-020-25538194,
email: amselvam@gmail.com

# A general systems theory for rain formation in warm clouds

## Abstract

A cumulus cloud model which can explain the observed characteristics of warm rain formation in monsoon clouds is presented. The model is based on classical statistical physical concepts and satisfies the principle of maximum entropy production. Atmospheric flows exhibit selfsimilar fractal fluctuations that are ubiquitous to all dynamical systems in nature, such as physical, chemical, social, etc and are characterized by inverse power law form for power (eddy energy) spectrum signifying long-range space-time correlations. A general systems theory model for atmospheric flows developed by the author is based on the concept that the large eddy energy is the integrated mean of enclosed turbulent (small scale) eddies. This model gives scale-free universal governing equations for cloud growth processes. The model predicted cloud parameters are in agreement with reported observations, in particular, the cloud drop-size distribution. Rain formation can occur in warm clouds within 30minutes lifetime under favourable conditions of moisture supply in the environment.

**1. Introduction**

Atmospheric flows exhibit self-similar fractal fluctuations generic to dynamical systems in nature. Self-similarity implies long-range space-time correlations identified as self-organized criticality (Bak et al. 1988). Yano et al. (2012) suggest that atmospheric convection could be an example of self–organized criticality. Atmospheric convection and precipitation have been hypothesized to be a real-world realization of self-organized criticality (SOC) (Peters et al. 2002, 2010). The physics of self-organized criticality ubiquitous to dynamical systems in nature and in finite precision computer realizations of non-linear numerical models of dynamical systems is not yet identified. Finite precision computer realizations of mathematical models (nonlinear) of dynamical systems do not give realistic solutions because of propagation of round-off error into mainstream computation (Selvam 1993; Sivak et al. 2013, Lawrence Berkeley National Laboratory 2013). During the past three decades, Lovejoy and his group (2010) have done extensive observational and theoretical studies of fractal nature of atmospheric flows and emphasize the urgent need to formulate and incorporate quantitative theoretical concepts of fractals in mainstream classical meteorological theory. The empirical analyses summarized by Lovejoy and Schertzer (2010) show that the statistical properties such as the mean and variance of atmospheric parameters (temperature, pressure, etc) are scale dependent and exhibit a power law relationship with a long fat tail over the space-time scales of measurement. The physics of the widely documented fractal fluctuations in dynamical systems is not yet identified. The traditional statistical normal (Gaussian) probability distribution is not applicable for statistical analysis of fractal space-time data sets because of the following reasons: (i) Gaussian distribution assumes independent (uncorrelated) data points while fractal fluctuations exhibit long-range correlations (ii) The probability distribution of fractal fluctuations exhibit a long fat tail, i.e., extreme events are of more common occurrence than given by the classical theory (Selvam 2013; Lovejoy and Schertzer 2010).

A general systems theory model for fractal fluctuations (Selvam 2013) predicts that the amplitude probability distribution as well as the power (variance) spectrum of fractal fluctuations follow the universal inverse power law $\tau^{-4\sigma}$ where $\tau$ is the golden mean ($\approx 1.618$) and $\sigma$ the normalized standard deviation. The atmospheric aerosol size spectrum is derived in terms of the universal inverse power law characterizing atmospheric eddy energy spectrum. A universal (scale independent) spectrum is derived for homogeneous (same density) suspended atmospheric particulate size distribution expressed as a function of the golden mean $\tau$ ($\approx 1.618$), the total number concentration and the mean volume radius (or diameter) of the particulate size spectrum. Knowledge of the mean volume radius and total number concentration is sufficient to compute the total particulate size spectrum at any location. In summary, the model predictions are (i) Fractal fluctuations can be resolved into an overall logarithmic spiral trajectory with the quasiperiodic Penrose tiling pattern for the internal structure. (ii) The probability distribution of fractal space-time fluctuations represents the power (variance) spectrum for fractal fluctuations and follows universal inverse power law form incorporating the golden mean. Such a result that the additive amplitudes of eddies when squared represent probability distribution is observed in the subatomic dynamics of quantum systems such as the electron or photon. Therefore the irregular or unpredictable fractal fluctuations exhibit quantumlike chaos. (iii) Atmospheric aerosols are held in suspension by the vertical velocity distribution (spectrum). The normalised atmospheric aerosol size spectrum is derived in terms of the universal inverse power law characterizing atmospheric eddy energy spectrum.

The complete theory relating to the formation of warm cumulus clouds and their responses to the hygroscopic particle seeding are presented. It is shown that warm rain formation can occur within a time period of 30 mins as observed in practice. Traditional cloud physical concepts for rain development requires over an hour for a full-sized raindrop to form (McGraw and Liu 2003).

*1.1 Current Concepts in Meteorological Theory and Limitations*

The nonequilibrium system of atmospheric flows is modelled with assumption of local thermodynamic equilibrium up to the stratopause at 50 km; molecular motion of atmospheric component gases is implicitly embodied in the gas constant (Tuck 2010). Nonequilibrium systems can be studied numerically, but despite decades of research, it is still very difficult to define the analytical functions from which to compute their statistics and have an intuition for how these systems behave (Parmeggiani 2012). Realistic mathematical modelling for simulation and prediction of atmospheric flows requires alternative theoretical concepts and analytical or error-free numerical computational techniques and therefore comes under the field of 'General Systems research' as explained in the following.

Space-time power law scaling and non-local connections exhibited by atmospheric flows have also been documented in other nonequilibrium dynamical systems, e.g. financial markets, neural network of brain, genetic networks, internet, road traffic, flocking behaviour of some animals and birds. Such universal behaviour has been subject of intensive study in recent years as 'complex systems' under the subject headings selforganised criticality, nonlinear dynamics and chaos, network theory, pattern formation, information theory, cybernetics (communication, control and adaptation). Complex system is a system composed of many interacting parts, such that the collective behaviour or "emergent" behaviours of those parts together is more than the sum of their individual behaviours (Newman 2011). Weather and climate are emergent properties of the complex adaptive system of atmospheric flows. Complex systems in different fields of study exhibit similar characteristics and therefore belong to the field of 'General Systems'. The terms 'general systems' and 'general systems research (or general systems theory) are due to Ludwig von Bertalanffy. According to Bertalanffy, general systems research is a discipline whose subject matter is "the formulation and derivation of those principles which are valid for 'systems' in general" (von Bertalanffy 1972; Klir 2001).

Skyttner (2005) quotes basic ideas of general systems theory formulated by Fredrich Hegel (1770-1831) as follows

i. The whole is more than the sum of the parts

ii. The whole defines the nature of the parts

iii. The parts cannot be understood by studying the whole

iv. The parts are dynamically interrelated or interdependent

In cybernetics, a system is maintained in dynamic equilibrium by means of communication and control between the constituent parts and also between the system and its environment (Skyttner 2005).

## 2. General systems theory for fractal space-time fluctuations in atmospheric flows

General systems theory for atmospheric flows is based on classical statistical physical concept where ensemble average represents the steady state values of parameters such as pressure, temperature etc of molecular systems (gases) independent of details of individual molecule. The ideas of statistical mechanics have been successfully extended to various disciplines to study complex systems (Haken 1977; Liu and Daum 2001). Liu and his group (Liu 1992, 1995; Liu et al. 1995; Liu and Hallett 1997, 1998) have applied the systems approach to study cloud droplet size distributions. Townsend (1956) had visualized large eddies as envelopes enclosing turbulent (smaller scale) eddies. General systems theory for atmospheric flows (Selvam 2013) visualizes the hierarchical growth of larger scale eddies from space-time integration of smaller scale eddies resulting in an atmospheric eddy continuum manifested in the selfsimilar fractal fluctuations of meteorological parameters. The basic thermodynamical parameters such as pressure, temperature, etc. are given by the same classical statistical physical formulae (kinetic theory of gases) for each component eddy (volume) of the atmospheric eddy continuum. It may be shown that the Boltzmann distribution for molecular energies also represents the eddy energy distribution in the atmospheric eddy continuum (Selvam 2013). In the following, general systems theory model concepts for atmospheric flows are summarized with model predictions for atmospheric flows and cloud growth parameters. Model predictions are compared with observations.

The atmospheric boundary layer (ABL), the layer extending up to about 2-3kms above the surface of the earth plays an important role in the formation of weather systems in the troposphere. It is important to identify and quantify the physical processes in the atmospheric boundary layer for realistic simulation of weather systems of all scales.

The ABL is often organized into helical secondary circulations which are often referred to as vortex roll or large eddies (Brown 1980). It is not known how these vortex rolls are sustained without decay by the turbulence around them. The author (Selvam 2013) has shown that the production of buoyant energy by the microscale fractional condensation (MFC) in turbulent eddies is responsible for the sustenance and growth of large eddies. Earlier Eady (1950) has emphasized the importance of large scale turbulence in the maintenance of the general circulation of the atmosphere.

The non-deterministic model described below incorporates the physics of the growth of macro-scale coherent structures from microscopic domain fluctuations in atmospheric flows (Selvam 2013). In summary, the mean flow at the planetary ABL possesses an inherent upward momentum flux of frictional origin at the planetary surface. This turbulence-scale upward momentum flux is progressively amplified by the exponential decrease of the atmospheric density with height coupled with the buoyant energy supply by micro-scale fractional condensation on hygroscopic nuclei, even in an unsaturated environment (Pruppacher and Klett 1997). The mean large-scale upward momentum flux generates helical vortex-roll (or large eddy) circulations in the planetary atmospheric boundary layer and is manifested as cloud rows and (or) streets, and meso-scale cloud clusters (MCC) in the global cloud cover pattern. A conceptual model (Selvam 2013) of large and turbulent eddies in the planetary ABL is shown in Figures. 1 and 2. The mean airflow at the planetary surface carries the signature of the fine scale features of the planetary surface topography as turbulent fluctuations with a net upward momentum flux. This persistent upward momentum flux of surface frictional origin generates large-eddy (or vortex-roll) circulations, which carry upward the turbulent eddies as internal circulations. Progressive upward growth of a large eddy occurs because of buoyant energy generation in turbulent fluctuations as a result of the latent heat of condensation of atmospheric water vapour on suspended hygroscopic nuclei such as common salt particles. The latent heat of condensation generated by the turbulent eddies forms a distinct warm envelope or a micro-scale capping inversion (MCI) layer at the crest of the large-eddy circulations as shown in Figure 1.

The turbulent eddies originating from surface friction exist all along the envelope of the large eddy (Figure 1) and the MFC takes place even in an unsaturated environment (Pruppachar and Klett 1997).

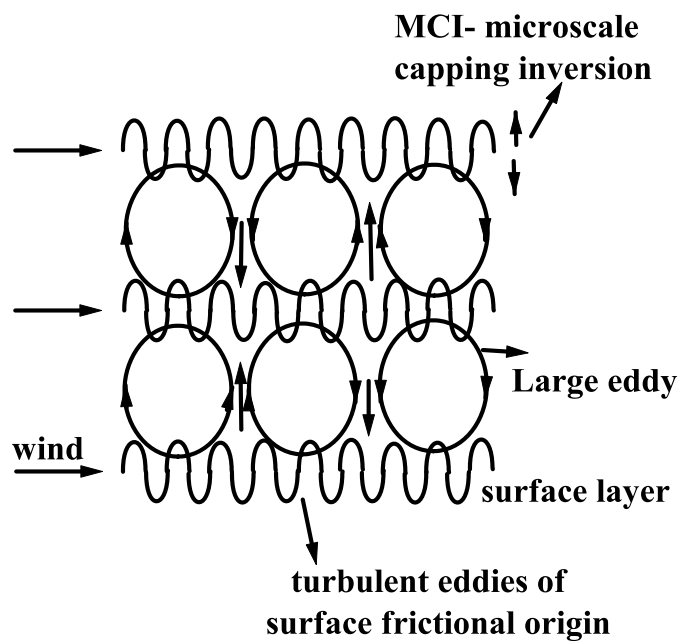

Figure 1. Eddies in the atmospheric planetary boundary layer

Progressive upward growth of the large eddy occurs from the turbulence scale at the planetary surface to a height $R$ and is seen as the rising inversion of the daytime atmospheric boundary layer (Figure 2).

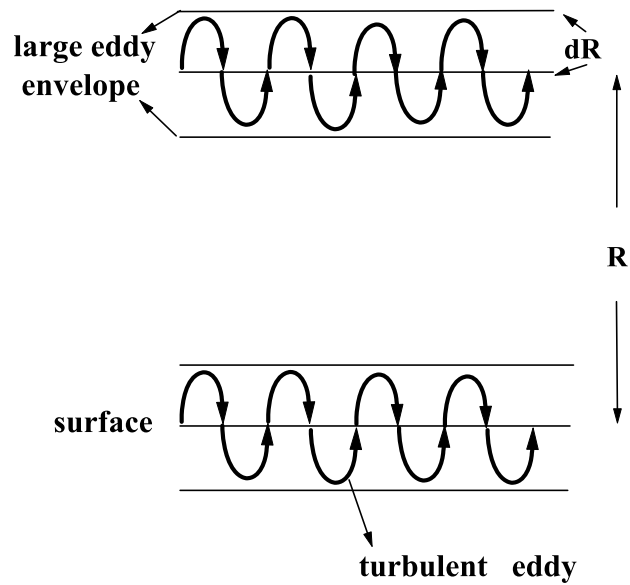

Figure 2. Growth of large eddy in the environment of the turbulent eddy

The turbulent fluctuations at the crest of the growing large-eddy mix overlying environmental air into the large-eddy volume, i.e. there is a two-stream flow of warm air upward and cold air downward analogous to superfluid turbulence in liquid helium (Donelly 1998, 1990). The convective growth of a large eddy in the atmospheric boundary layer therefore occurs by vigorous counter flow of air in turbulent fluctuations, which releases stored buoyant energy in the medium of propagation, e.g., latent heat of condensation of atmospheric water vapour. Such a picture of atmospheric convection is different from the traditional concept of atmospheric eddy growth by diffusion, i.e. analogous to the molecular level momentum transfer by collision (Selvam 2013).

The generation of turbulent buoyant energy by the micro-scale fractional condensation is maximum at the crest of the large eddies and results in the warming of the large-eddy volume. The turbulent eddies at the crest of the large eddies are identifiable by a micro-scale capping inversion that rises upward with the convective growth of the large eddy during the course of the day. This is seen as the rising inversion of the daytime planetary boundary layer in echosonde and radiosonde records and has been identified as the entrainment zone (Boers 1989; Gryning and Batchvarova 2006) where mixing with the environment occurs.

In summary, the atmospheric boundary layer (ABL) contains large eddies (vortex rolls) which carry on their envelopes turbulent eddies of surface frictional origin (Selvam 2013). The buoyant energy production by *microscale-fractional-condensation* (MFC) in turbulent eddies is responsible for the sustenance and growth of large eddies.

The buoyant energy production of turbulent eddies by the MFC process is maximum at the crest of the large eddies and results in the warming of the large eddy volume. The turbulent eddies at the crest of the large eddies are identifiable by a *microscale-capping-inversion* (MCI) layer which rises upwards with the convective growth of the large eddy in the course of the day. The MCI layer is a region of enhanced aerosol concentrations. The atmosphere contains a stack of large eddies. Vertical mixing of overlying environmental air into the large eddy volume occurs by turbulent eddy fluctuations (Selvam 2013). The energy gained by the turbulent eddies would contribute to the sustenance and growth of the large eddy.

## *2.1 Large eddy growth in the atmospheric boundary layer*

Townsend (1956) has visualized the large eddy as the integrated mean of enclosed turbulent eddies. The r.m.s (root mean square) circulation speed $W$ of the large eddy of radius $R$ is expressed in terms of the enclosed turbulent eddy circulation speed $w_*$ and radius $r$ as (Selvam 2013).

$$W^2 = \frac{2}{\pi}\frac{r}{R} w_*^2 \qquad (1)$$

Based on Townsend's concept, the observed eddy continuum eddy growth in the atmospheric boundary layer is visualized to occur in the following two stages (see Section 2.2 below). (i) Growth of primary dominant turbulent eddy circulation from successive equal fluctuation length step increment d$z$ equal to one. Identifiable organized whole turbulent eddy circulation forms at length step $z = 10$ associated with fractional volume dilution by eddy mixing less than half so that its (eddy) identity is not erased. (ii) Large eddies are then visualized to form as envelopes enclosing these dominant turbulent eddies starting from unit primary eddy as zero level. Spatial domain of large eddy is obtained by integration from initial radius $r$ to large eddy radius $R$ (Eq. 1). The growing large eddy traces logarithmic spiral circulation with quasiperiodic Penrose tiling pattern for the internal structure such that successive large eddy radii follow the Fibonacci number series and ratio of successive radii $R_{n+1}/R_n$ is equal to the golden mean $\tau$ ( $\approx 1.618$). Further it is shown that the variance (square of amplitude) spectrum and amplitude probability distribution of fractal fluctuations are represented by the same function, i.e., $W$ and $W^2$ have the same probability distribution. Such a result that the additive amplitude of eddies when squared represent the probability distribution of amplitudes is exhibited by microscopic scale quantum systems such as the electron or photon. Fractal fluctuations therefore exhibit quantum-like chaos. At each level the integrated mean of eddy circulation speed $w_*$ gives the mean circulation speed $W$ for the next level. Therefore $W$ and $w_*$ represent respectively the mean and corresponding standard deviation of eddy circulation speed at that level and the ratio $W/w_*$ is the normalized standard deviation σ. The primary turbulent eddy with circulation speed $w_*$ and radius $r$ is the reference level and normalized standard deviation σ values from 0 to 1 refer to the primary eddy growth region. Primary eddy growth begins with unit length step perturbation followed by successive 10 unit length growth steps (Selvam 2013; see Section 3.3 below).

*2.2 Primary dominant eddy growth mechanism in the ABL*

*2.2.1 Steady state fractional volume dilution of large eddy by turbulent fluctuations*

As seen from Figures 1 and 2 and from the concept of large eddy growth, vigorous counter flow (mixing) in turbulent eddy fluctuations characterizes the large-eddy volume. The total fractional volume dilution rate of the large eddy by turbulent (eddy) vertical mixing of environmental air across unit cross-section of the large eddy surface is derived from (1) and is given as follows.

The ratio of the upward mass flux of air in the turbulent eddy to that in the large eddy across unit cross-section (of the large eddy) per second is equal to $w_*/dW$ where $w_*$ is the increase in vertical velocity per second of the turbulent eddy due to the microscale fractional condensation (MFC) process, and $dW$ is the corresponding increase in vertical velocity of large eddy. This fractional volume dilution of the large eddy occurs in the environment of the turbulent eddy. The fractional volume of the large eddy which is in the environment of the turbulent eddy where dilution occurs is equal to $r/R$.

Therefore, the total fractional volume dilution $k$ of the large eddy per second across unit cross-section can be expressed as

$$k = \frac{w_*}{\mathrm{d}W} \frac{r}{R} \qquad (2)$$

The value of $k \approx 0.4$ when the length scale ratio $R/r$ is equal to 10 since $\mathrm{d}W \approx 0.25\ w_*$ (1). The growing large eddy cannot exist as a recognizable entity for length scale ratio values less than 10.

Identifiable large eddies can grow only for scale ratios $z > 10$. The convective scale eddy of radius $R_c$ evolves from the turbulent eddy of radius $r$ for the size ratio ($z$). $R_c/r = 10$. This type of decadic scale range eddy mixing can be visualized to occur in successive decadic scale ranges generating the convective, meso-, synoptic and planetary scale eddies of radii $R_c$, $R_m$, $R_s$, and $R_p$ where c, m, s and p represent respectively the convective, meso-, synoptic and planetary scales.

### 2.2.2 Logarithmic wind profile in the ABL

The height interval in which this incremental change $\mathrm{d}W$ in the vertical velocity occurs is $\mathrm{d}R$ which is equal to $r$. The height up to which the large eddy has grown is equal to $R$ (see Figure 2).

Using the above expressions (2) can be written as

$$\mathrm{d}W = \frac{w_*}{k} \frac{\mathrm{d}R}{R} = \frac{w_*}{k} \mathrm{d}\ln R \qquad (3)$$

Integrating (3) between the height interval $r$ and $R$ the following relation for $W$ can be obtained as

$$W = \int_r^R \frac{w_*}{k} \mathrm{d}\ln R = \frac{w_*}{k}(\ln R - \ln r)$$
$$W = \frac{w_*}{k} \ln\left(\frac{R}{r}\right) = \frac{w_*}{k} \ln z \qquad (4)$$

In (4) it is assumed that $w_*$ and $r_*$ are constant for the height interval of integration. The length scale ratio $R/r$ is denoted by $z$. A normalized height with reference to the turbulence scale $r$ can be defined as $z = R/r$.

Equation (4) is the well known logarithmic velocity profiles in turbulent shear flows discussed originally by von Karman (1930) and Prandtl (1932) (for a recent review, see Marusic et al. 2010), and to the recently discovered logarithmic variation of turbulent fluctuations in pipe flow (Hultmark et al. 2012). Observations and the existing theory of eddy diffusion (Holton 2004) indicate that the vertical wind profile in the ABL follows the logarithmic law which is identical to the expression shown in (4). The constant $k$ (Von Karman's constant) as determined from observations is equal to 0.4 and has not been assigned any physical meaning in the literature. The new theory proposed in the present study enables prediction of observed logarithmic wind profile without involving any assumptions as in the case of existing theories of atmospheric diffusion processes such as molecular momentum transfer (Holton 2004). Also, the constant $k$ now has a physical meaning, namely, it is the fractional volume dilution rate of the large eddy by the turbulent scale eddies for dominant large eddy growth.

*2.2.3 Fractional upward mass flux of surface air*

Vertical mixing due to turbulent eddy fluctuations progressively dilutes the rising large eddy and a fraction $f$ equal to $Wr/w_*R$ of surface air reaches the normalised height $z$ as shown in the following. The turbulent eddy fluctuations carry upward surface air of frictional origin. The ratio of upward mass flux of air /unit time/unit cross-section in the large eddy to that in the turbulent eddy $= W/w_*$. The magnitude of $W$ is smaller than that of $w_*$ (1). The ratio $W/w_*$ is equal to the upward mass flux of surface air /unit time/unit cross-section in the rising large eddy. The volume fraction of turbulent eddy across unit cross-section of large eddy envelope is equal to $r/R$. Turbulence scale upward mass flux of surface air equal to $W/w_*$ occurs in the fractional volume $r/R$ of the large eddy. Therefore the net upward mass flux $f$ of surface air/unit time/unit cross-section in the large eddy environment is equal to

$$f = \frac{W}{w_*} \frac{r}{R} \tag{5}$$

The large eddy circulation speed $W$ and the corresponding temperature perturbation $\theta$ may be expressed in terms of $f$ and $z$ as

$$\begin{aligned} W &= w_* fz \\ \theta &= \theta_* fz \end{aligned} \tag{6}$$

In (6) $\theta_*$ is the temperature perturbation corresponding to the primary turbulent eddy circulation speed $w_*$.

The corresponding moisture content $q$ at height $z$ is related to the moisture content $q_*$ at the surface and is given as

$$q = q_* fz \tag{7}$$

Substituting from (1), (2) and (4), the net upward flux $f$ of surface air at level $z$ is now obtained from (5) as

$$f = \frac{1}{kz} \ln z = \frac{WR}{w_* rz} \ln z = \sqrt{\frac{2}{\pi z}} \ln z \tag{8}$$

In (8) $f$ represents the steady state fractional volume of surface air at any level $z$. A fraction $f$ of surface aerosol concentration $N_*$ is carried upward to normalised height $z$. The aerosol number concentration $N$ at level $z$ is then given as

$$N = N_* f \tag{9}$$

Since atmospheric aerosols originate from surface, the vertical profile of mass and number concentration of aerosols follow the $f$ distribution. The vertical mass exchange mechanism predicts the $f$ distribution for the steady state vertical transport of aerosols at higher levels. The vertical variation of atmospheric aerosol number concentration given by the $f$ distribution is shown in Figure 3. The vertical variation of large eddy circulation speed $W$ and the corresponding temperature $\theta$ are shown respectively in Figures 4 and 5.

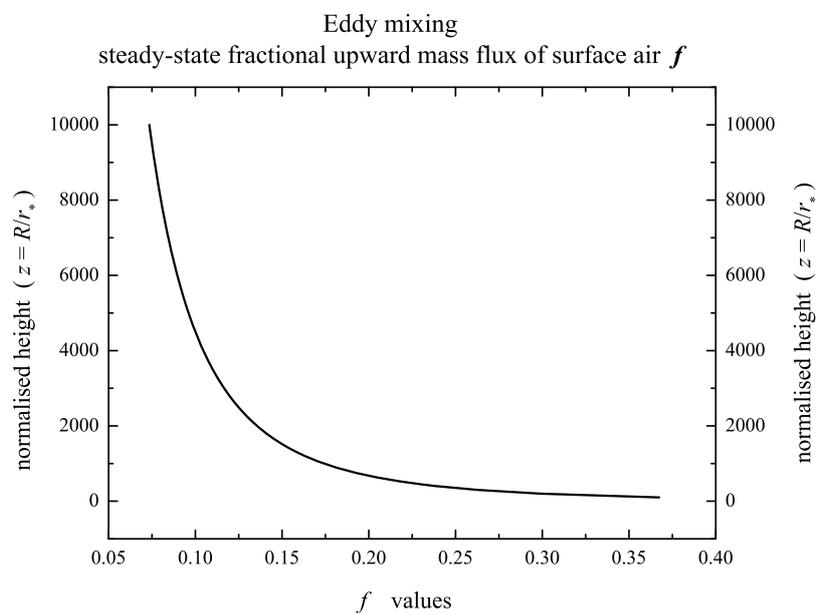

Figure 3. $f$ distribution for $r_* =1$. The above $f$ distribution represents vertical distribution of (i) atmospheric aerosol concentration (ii) ratio of cloud liquid water content to adiabatic liquid water content

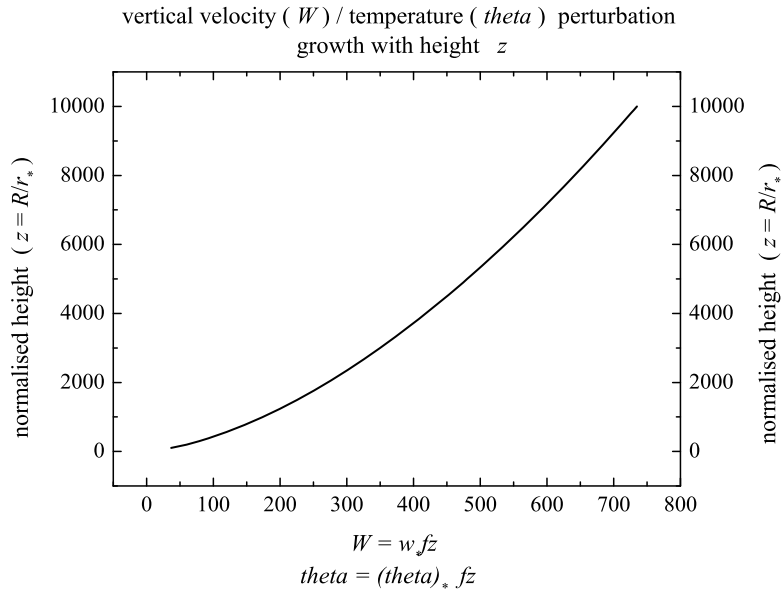

Figure 4. Vertical variation of large eddy circulation speed $W$ and temperature $\theta$

## 3. Atmospheric aerosol (particulates) size spectrum

The atmospheric eddies hold in suspension the aerosols and thus the size spectrum of the atmospheric aerosols is dependent on the vertical velocity spectrum of the atmospheric eddies as shown below. Earlier Liu (1956) has studied the problem of the dispersion of material particles in a turbulent fluid and remarks that particle dispersion constitutes a direct and striking manifestation of the mechanism of fluid turbulence. Grabowski and Wang (2012) discuss multiscale nature of turbulent cloud microphysical processes and its significant impact on warm rain initiation.

The aerosols are held in suspension by the eddy vertical velocity perturbations. Thus the suspended aerosol mass concentration $m$ at any level $z$ will be directly related to the vertical velocity perturbation $W$ at $z$, i.e., $W \sim mg$ where $g$ is the acceleration due to gravity. Substituting in (6) for $W$ and $w_*$ in terms of aerosol mass concentrations $m$ and $m_*$ respectively at normalized height $z$ and at surface layer, the vertical variation of aerosol mass concentration flux is obtained as

$$m = m_* fz \tag{10}$$

*3.1 Vertical variation of aerosol mean volume radius*

The mean volume radius of aerosol increases with height as shown in the following.

The velocity perturbation $W$ is represented by an eddy continuum of corresponding size (length) scales $z$. The aerosol mass flux across unit cross-section per unit time is obtained by normalizing the velocity perturbation $W$ with respect to the corresponding length scale $z$ to give the volume flux of air equal to $Wz$ and can be expressed as follows from (6):

$$Wz = (w_* fz)z = w_* fz^2 \tag{11}$$

The corresponding normalised moisture flux perturbation is equal to $qz$ where $q$ is the moisture content per unit volume at level $z$. Substituting for $q$ from (7)

$$qz = \text{normalised moisture flux at level } z = q_* fz^2 \tag{12}$$

The moisture flux increases with height resulting in increase of mean volume radius of cloud condensation nuclei (CCN) because of condensation of water vapour. The corresponding CCN (aerosol) mean volume radius $r_a$ at height $z$ is given in terms of the aerosol number concentration $N$ at level $z$ and mean volume radius $r_{as}$ at the surface as follows from (12)

$$\frac{4}{3}\pi r_a^3 N = \frac{4}{3}\pi r_{as}^3 N_* fz^2 \tag{13}$$

Substituting for $N$ from (9) in terms of $N_*$ and $f$

$$\begin{aligned} r_a^3 &= r_{as}^3 z^2 \\ r_a &= r_{as} z^{2/3} \end{aligned} \tag{14}$$

The mean aerosol size increases with height according to the cube root of $z^2$ (14). As the large eddy grows in the vertical, the aerosol size spectrum extends towards larger sizes while the total number concentration decreases with height according to the $f$ distribution.

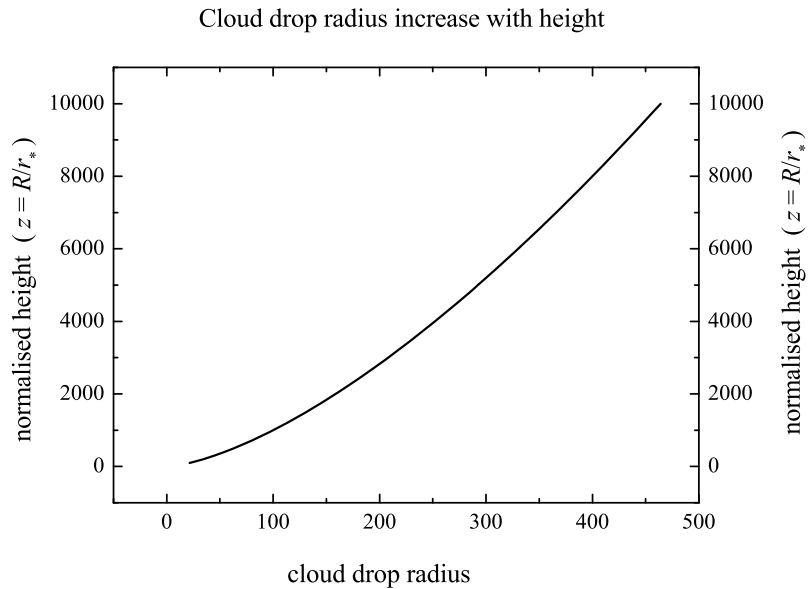

Figure 5. Vertical variation of aerosol mean volume radius with height

The atmospheric aerosol size spectrum is dependent on the eddy energy spectrum and may be expressed in terms of the recently identified universal characteristics of fractal fluctuations generic to atmospheric flows (Selvam 2013) as shown in Section 3.2 below.

*3.2 Probability distribution of fractal fluctuations in atmospheric flows*

The atmospheric eddies hold in suspension atmospheric particulates, namely, aerosols, cloud drops and raindrops and the size spectrum of these atmospheric suspended particulates is dependent on the vertical velocity spectrum of the atmospheric eddies. Atmospheric air flow is turbulent, i.e., consists of irregular fluctuations of all space-time scales characterized by a broadband spectrum of eddies. The suspended particulates will also exhibit a broadband size spectrum closely related to the atmospheric eddy energy spectrum.

It is now established (Lovejoy and Schertzer 2010) that atmospheric flows exhibit self-similar fractal fluctuations generic to dynamical systems in nature such as fluid flows, heart beat patterns, population dynamics, spread of forest fires, etc. Power spectra of fractal fluctuations exhibit inverse power law of form - $f^\alpha$ where α is a constant indicating long-range space-time correlations or persistence. Inverse power law for power spectrum indicates scale invariance, i.e., the eddy energies at two different scales (space-time) are related to each other by a scale factor (α in this case) alone independent of the intrinsic properties such as physical, chemical, electrical etc of the dynamical system.

A general systems theory for turbulent fluid flows predicts that the eddy energy spectrum, i.e., the variance (square of eddy amplitude) spectrum is the same as the probability distribution $P$ of the eddy amplitudes, i.e. the vertical velocity $W$ values. Such a result that the additive amplitudes of eddies, when squared, represent the probabilities is exhibited by the subatomic dynamics of quantum systems such as the electron or photon. Therefore the unpredictable or irregular fractal space-time fluctuations generic to dynamical systems in nature, such as atmospheric flows is a signature of quantum-like chaos. The general systems theory for turbulent fluid flows predicts (Selvam 2013) that the atmospheric eddy energy spectrum represented by the probability distribution $P$ follows inverse power law form incorporating the *golden mean* τ and the normalized deviation σ for values of σ ≥ 1 and σ ≤ -1 as given below

$$P = \tau^{-4\sigma} \qquad (15)$$

The vertical velocity *W* spectrum will therefore be represented by the probability distribution *P* for values of σ ≥ 1 and σ ≤ -1 given in (15) since fractal fluctuations exhibit quantum-like chaos as explained above.

$$W = P = \tau^{-4\sigma} \quad (16)$$

Values of the normalized deviation σ in the range -1 < σ < 1 refer to regions of primary eddy growth where the fractional volume dilution *k* (2) by eddy mixing process has to be taken into account for determining the probability distribution *P* of fractal fluctuations (see Section 3.3 below).

### *3.3 Primary eddy growth region fractal fluctuation probability distribution*

Normalised deviation σ ranging from -1 to +1 corresponds to the primary eddy growth region. In this region the probability *P* is shown to be equal to $P = \tau^{-4k}$ (see below) where *k* is the fractional volume dilution by eddy mixing (2).

The normalized deviation σ represents the length step growth number for growth stages more than one. The first stage of eddy growth is the primary eddy growth starting from unit length scale (*r* = 1) perturbation, the complete eddy forming at the tenth length scale growth, i.e., *R* = 10*r* and scale ratio *z* equals 10 (Selvam 2013). The steady state fractional volume dilution *k* of the growing primary eddy by internal smaller scale eddy mixing is given by (2) as

$$k = \frac{w_* r}{WR} \quad (17)$$

The expression for *k* in terms of the length scale ratio *z* equal to *R*/*r* is obtained from (1) as

$$k = \sqrt{\frac{\pi}{2z}} \qquad (18)$$

A fully formed large eddy length $R = 10r$ ($z=10$) represents the average or mean level zero and corresponds to a maximum of 50% cumulative probability of occurrence of either positive or negative fluctuation peak at normalized deviation σ value equal to zero by convention. For intermediate eddy growth stages, i.e., $z$ less than 10, the probability of occurrence of the primary eddy fluctuation does not follow conventional statistics, but is computed as follows taking into consideration the fractional volume dilution of the primary eddy by internal turbulent eddy fluctuations. Starting from unit length scale fluctuation, the large eddy formation is completed after 10 unit length step growths, i.e., a total of 11 length steps including the initial unit perturbation. At the second step ($z = 2$) of eddy growth the value of normalized deviation σ is equal to 1.1 - 0.2 (= 0.9) since the complete primary eddy length plus the first length step is equal to 1.1. The probability of occurrence of the primary eddy perturbation at this σ value however, is determined by the fractional volume dilution $k$ which quantifies the departure of the primary eddy from its undiluted average condition and therefore represents the normalized deviation σ. Therefore the probability density $P$ of fractal fluctuations of the primary eddy is given using the computed value of $k$ as shown in the following equation.

$$P = \tau^{-4k} \qquad (19)$$

The vertical velocity $W$ spectrum will therefore be represented by the probability density distribution $P$ for values of $-1 \leq \sigma \leq 1$ given in (19) since fractal fluctuations exhibit quantum-like chaos as explained above (16).

$$W = P = \tau^{-4k} \qquad (20)$$

The probabilities of occurrence ($P$) of the primary eddy for a complete eddy cycle either in the positive or negative direction starting from the peak value ($\sigma = 0$) are given for progressive growth stages ($\sigma$ values) in the following Table 1. The statistical normal probability density distribution corresponding to the normalized deviation $\sigma$ values are also given in the Table 1.

Table 1: Primary eddy growth

| Growth step no | ± σ | k | Probability (%) | |
|---|---|---|---|---|
| | | | Model predicted | Statistical normal |
| 2 | .9000 | .8864 | 18.1555 | 18.4060 |
| 3 | .8000 | .7237 | 24.8304 | 21.1855 |
| 4 | .7000 | .6268 | 29.9254 | 24.1964 |
| 5 | .6000 | .5606 | 33.9904 | 27.4253 |
| 6 | .5000 | .5118 | 37.3412 | 30.8538 |
| 7 | .4000 | .4738 | 40.1720 | 34.4578 |
| 8 | .3000 | .4432 | 42.6093 | 38.2089 |
| 9 | .2000 | .4179 | 44.7397 | 42.0740 |
| 10 | .1000 | .3964 | 46.6250 | 46.0172 |
| 11 | 0 | .3780 | 48.3104 | 50.0000 |

The model predicted probability density distribution $P$ along with the corresponding statistical normal distribution with probability values plotted on linear and logarithmic scales respectively on the left and right hand sides are shown in Figure 6. The model predicted probability distribution $P$ for fractal space-time fluctuations is very close to the statistical normal distribution for normalized deviation $\sigma$ values less than 2 as seen on the left hand side of Figure 6. The model predicts progressively higher values of probability $P$ for values of $\sigma$ greater than 2 as seen on a logarithmic plot on the right hand side of Figure 6.

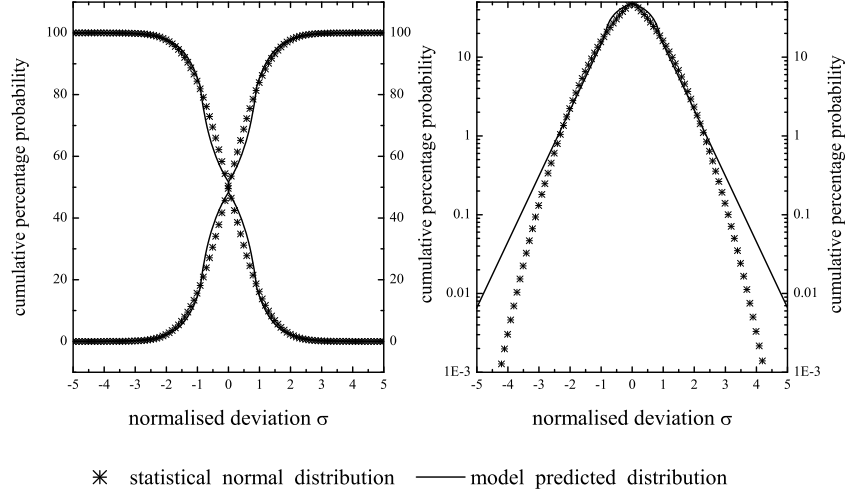

fractal fluctuations probability distribution
comparison with statistical normal distribution

✳ statistical normal distribution ——— model predicted distribution

Figure 6: Model predicted probability distribution $P$ along with the corresponding statistical normal distribution with probability values plotted on linear and logarithmic scales respectively on the left and right hand side graphs.

### 3.4 Atmospheric wind spectrum and suspended particulate size spectrum

The steady state flux $dN$ of CCN at level $z$ in the normalised vertical velocity perturbation $(dW)z$ is given as

$$dN = N(dW)z \qquad (21)$$

The logarithmic wind profile relationship for $W$ at (4) gives

$$dN = Nz\frac{w_*}{k}d(\ln z) \qquad (22)$$

Substituting for $k$ from (2)

$$dN = Nz\frac{w_*}{w_*}Wzd(\ln z) = NWz^2 d(\ln z) \qquad (23)$$

The length scale $z$ is related to the suspended particulate radius $r_a$ (14).

Therefore

$$\ln z = \frac{3}{2} \ln\left(\frac{r_a}{r_{as}}\right) \qquad (24)$$

Defining a normalized radius $r_{an}$ equal to $\frac{r_a}{r_{as}}$, i.e., $r_{an}$ represents the suspended particulate mean volume radius $r_a$ in terms of its mean volume radius $r_{as}$ at the surface (or reference level). Therefore

$$\ln z = \frac{3}{2} \ln r_{an} \qquad (25)$$

$$d \ln z = \frac{3}{2} d \ln r_{an} \qquad (26)$$

Substituting for dln$z$ in (23)

$$dN = NWz^2 \frac{3}{2} d(\ln r_{an}) \qquad (27)$$

$$\frac{dN}{d(\ln r_{an})} = \frac{3}{2} NWz^2 \qquad (28)$$

The above equation is for the scale length $z$. The volume across unit cross-section associated with scale length $z$ is equal to $z$. The particle radius corresponding to this volume is equal to $z^{1/3}$

The above equation 28 is for the scale length $z$ and the corresponding radius equal to $z^{1/3}$. The equation (28) normalised for scale length and associated drop radius is given as

$$\frac{dN}{d(\ln r_{an})} = \frac{3}{2} \frac{NPz^2}{z \times z^{\frac{1}{3}}} = \frac{3}{2} NWz^{\frac{2}{3}} \qquad (28a)$$

Substituting for $W$ from (16) and (20) in terms of the universal probability density $P$ for fractal fluctuations

$$\frac{dN}{d(\ln r_{an})} = \frac{3}{2} NP z^{\frac{2}{3}} \quad (29)$$

The general systems theory predicts that fractal fluctuations may be resolved into an overall logarithmic spiral trajectory with the quasiperiodic Penrose tiling pattern for the internal structure such that the successive eddy lengths follow the Fibonacci mathematical number series (Selvam 2013). The eddy length scale ratio $z$ for length step $\sigma$ is therefore a function of the golden mean $\tau$ given as

$$z = \tau^{\sigma} \quad (30)$$

Expressing the scale length $z$ in terms of the golden mean $\tau$ in (29)

$$\frac{dN}{d(\ln r_{an})} = \frac{3}{2} NP \tau^{\frac{2\sigma}{3}} \quad (31)$$

In (31) $N$ is the steady state aerosol concentration at level $z$. The normalized aerosol concentration spectrum any level $z$ is given as

$$\frac{1}{N} \frac{dN}{d(\ln r_{an})} = \frac{3}{2} P \tau^{\frac{2\sigma}{3}} \quad (32)$$

The fractal fluctuations probability density is $P = \tau^{-4\sigma}$ (16) for values of the normalized deviation $\sigma \geq 1$ and $\sigma \leq -1$ on either side of $\sigma = 0$ as explained earlier (Sections 3.3, 3.4). Values of the normalized deviation $-1 \leq \sigma \leq 1$ refer to regions of primary eddy growth where the fractional volume dilution $k$ (2) by eddy mixing process has to be taken into account for determining the probability density $P$ of fractal fluctuations. Therefore the probability density $P$ in the primary eddy growth region ($\sigma \geq 1$ and $\sigma \leq -1$) is given using the computed value of $k$ as $P = \tau^{-4k}$ (20).

The normalised radius $r_{an}$ is given in terms of $\sigma$ and the golden mean $\tau$ from (25) and (30) as follows.

$$\ln z = \frac{3}{2} \ln r_{an}$$
$$r_{an} = z^{2/3} = \tau^{2\sigma/3}$$
(33)

The normalized aerosol size spectrum is obtained by plotting a graph of normalized aerosol concentration $\frac{1}{N}\frac{dN}{d(\ln r_{an})} = \frac{3}{2}P\tau^{\frac{2\sigma}{3}}$ (32) versus the normalized aerosol radius $r_{an} = \tau^{2\sigma/3}$ (33). The normalized aerosol size spectrum is derived directly from the universal probability density $P$ distribution characteristics of fractal fluctuations (16 and 20) and is independent of the height $z$ of measurement and is universal for aerosols in turbulent atmospheric flows. The aerosol size spectrum is computed starting from the minimum size, the corresponding probability density $P$ (32) refers to the cumulative probability density starting from 1 and is computed as equal to $P = 1 - \tau^{-4\sigma}$.

The universal normalised aerosol size spectrum represented by $\frac{1}{N}\frac{dN}{d(\ln r_{an})}$ versus $r_{an}$ is shown in Figure 7.

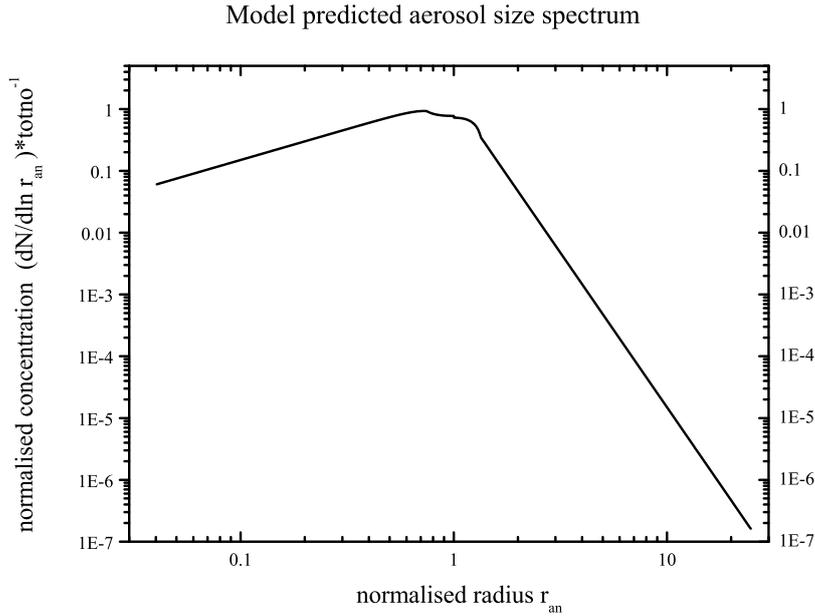

Figure 7: Model predicted aerosol size spectrum

## 3.5 Large eddy growth time

The time $\Gamma$ taken for the steady state aerosol concentration $f$ to be established at the normalised height $z$ is equal to the time taken for the large eddy to grow to the height $z$ and is computed as follows.

The time required for the large eddy of radius $R$ to grow from the primary turbulence scale radius $r_*$ is computed as follows.

$$\text{The scale ratio } z = \frac{R}{r_*}$$

Therefore for constant turbulence radius $r_*$

$$dz = \frac{dR}{r_*}$$

The incremental growth $dR$ of large eddy radius is equal to

$$dR = r_* dz$$

The time period d$t$ for the incremental cloud growth is expressed as follows

$$dt = \frac{dR}{W} = \frac{r_* dz}{W}$$

The large eddy circulation speed $W$ is expressed in terms of $f$ and $z$ as (6)

$$W = w_* fz$$

Substituting for $f$ from (8)

$$W = w_* z \sqrt{\frac{2}{\pi z}} \ln z = w_* \sqrt{\frac{2z}{\pi}} \ln z$$

The time $\Gamma$ taken for large eddy growth from surface to normalized height $z$ is obtained as

$$\Gamma = \int dt = \frac{r_*}{w_*} \sqrt{\frac{\pi}{2}} \int_2^z \frac{dz}{z^{1/2} \ln z}$$

The above equation can be written in terms of $\sqrt{z}$ as follows

$$d(z^{0.5}) = \frac{dz}{2\sqrt{z}}$$
$$dz = 2\sqrt{z}\, d(\sqrt{z})$$

Therefore substituting in (35)

$$\Gamma = \frac{r_*}{w_*} \sqrt{\frac{\pi}{2}} \int_2^z \frac{2\sqrt{z}\, d\sqrt{z}}{z^{1/2} \ln z} = \frac{r_*}{w_*} \sqrt{\frac{\pi}{2}} \int_2^z \frac{d\sqrt{z}}{\left(\frac{1}{2}\ln z\right)}$$

$$\Gamma = \frac{r_*}{w_*}\sqrt{\frac{\pi}{2}}\int_{x1}^{x2}\frac{d(\sqrt{z})}{\ln\sqrt{z}} = \frac{r_*}{w_*}\sqrt{\frac{\pi}{2}}\int_{x1}^{x2}\mathrm{li}(\sqrt{z})$$

$$x_1 = \sqrt{z_1} \text{ and } x_2 = \sqrt{z_2}$$

In the above equation $z_1$ and $z_2$ refer respectively to lower and upper limits of integration and li is the Soldner's integral or the logarithm integral. The large eddy growth time $\Gamma$ can be computed from (36).

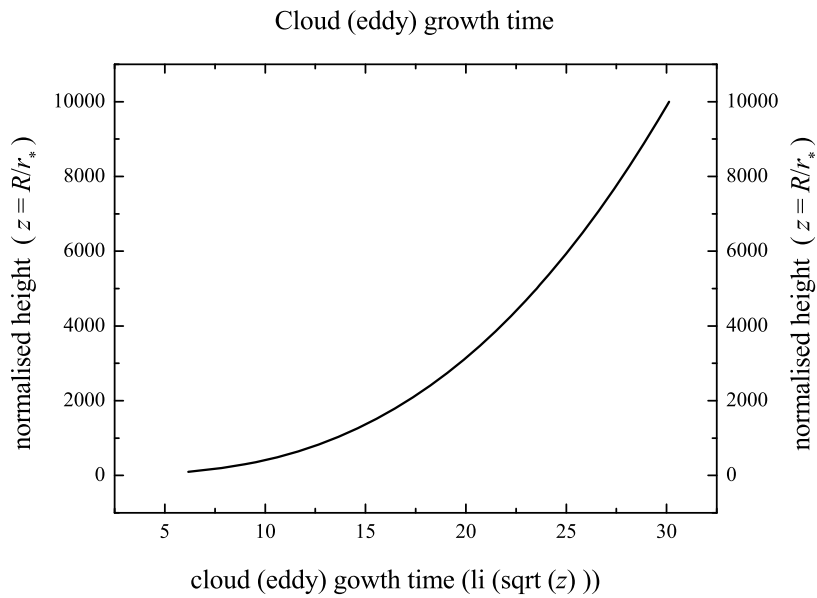

Figure 8. Large eddy (cloud) growth time

**4. General systems theory and maximum entropy principle**

The maximum entropy principle concept of classical statistical physics is applied to determine the fidelity of the inverse power law probability distribution $P$ (16 and 20) for exact quantification of the observed space-time fractal fluctuations of dynamical systems ranging from the microscopic dynamics of quantum systems to macro-scale real world dynamical systems. Kaniadakis (2009) states that the correctness of an analytic expression for a given power-law tailed distribution used to describe a statistical system is strongly related to the validity of the generating mechanism. In this sense the maximum entropy principle, the cornerstone of statistical physics, is a valid and powerful tool to explore new roots in searching for generalized statistical theories (Kaniadakis 2009). The concept of entropy is fundamental in the foundation of statistical physics. It first appeared in thermodynamics through the second law of thermodynamics. In statistical mechanics, we are interested in the disorder in the distribution of the system over the permissible microstates. The measure of disorder first provided by Boltzmann principle (known as Boltzmann entropy) is given by $S = K_B \ln M$, where $K_B$ is the thermodynamic unit of measurement of entropy and is known as Boltzmann constant. $K_B = 1.38 \times 10^{-16}$ erg/°C. The variable $M$, called thermodynamic probability or statistical weight, is the total number of microscopic complexions compatible with the macroscopic state of the system and corresponds to the "degree of disorder" or 'missing information' (Chakrabarti and De 2000). For a probability distribution among a discrete set of states the generalized entropy for a system out of equilibrium is given as (Salingaros and West 1999; Chakrabarti and De 2000; Beck 2009; Sethna 2009).

$$S = -\sum_{j=1}^{\sigma} P_j \ln P_j$$

In (37) $P_j$ is the probability for the $j^{th}$ stage of eddy growth in the present study, $\sigma$ is the length step growth which is equal to the normalized deviation and the entropy $S$ represents the 'missing information' regarding the probabilities. Maximum entropy $S$ signifies minimum preferred states associated with scale-free probabilities.

The validity of the probability distribution $P$ (16 and 20) is now checked by applying the concept of maximum entropy principle (Kaniadakis 2009). The probability distribution $P$ is equal to -log$P$ as shown in the following

The r.m.s circulation speed $W$ of the large eddy follows a logarithmic relationship with respect to the length scale ratio $z$ equal to $R/r$ (3) as given below

$$W = \frac{w_*}{k} \log z$$

In the above equation the variable $k$ represents for each step of eddy growth, the fractional volume dilution of large eddy by turbulent eddy fluctuations carried on the large eddy envelope (Selvam 2013) and is given as (17)

$$k = \frac{w_* r}{WR}$$

Substituting for $k$ in (3) we have

$$W = w_* \frac{WR}{w_* r} \log z = \frac{WR}{r} \log z$$
and
$$\frac{r}{R} = \log z \qquad (38)$$

The ratio $r/R$ represents the fractional probability $P$ of occurrence of small-scale fluctuations ($r$) in the large eddy ($R$) environment. Since the scale ratio $z$ is equal to $R/r$, (38) may be written in terms of the probability $P$ as follows.

$$\frac{r}{R} = \log z = \log\left(\frac{R}{r}\right) = \log\left(\frac{1}{(r/R)}\right)$$
$$P = \log\left(\frac{1}{P}\right) = -\log P \qquad (39)$$

Substituting for log $P_j$ (39) and for the probability $P_j$ in terms of the golden mean $\tau$ derived earlier (16 and 20) the entropy $S$ (37) is expressed as

$$S = -\sum_{j=1}^{\sigma} P_j \log P_j = \sum_{j=1}^{\sigma} P_j^2 = \sum_{j=1}^{\sigma} \left(\tau^{-4\sigma}\right)^2$$
$$S = \sum_{j=1}^{\sigma} \tau^{-8\sigma} \approx 1 \text{ for large } \sigma \qquad (40)$$

In (40) $S$ is equal to the square of the cumulative probability density distribution and it increases with increase in $\sigma$, i.e., the progressive growth of the eddy continuum and approaches 1 for large $\sigma$. According to the second law of thermodynamics, increase in entropy signifies approach of dynamic equilibrium conditions with scale-free characteristic of fractal fluctuations and hence the probability distribution $P$ (16 and 20) is the correct analytic expression quantifying the eddy growth processes visualized in the general systems theory. The ordered growth of the atmospheric eddy continuum is associated with maximum entropy production.

Paltridge (2009) states that the principle of maximum entropy production (MEP) is the subject of considerable academic study, but is yet to become remarkable for its practical applications. The ability of a system to dissipate energy and to produce entropy "ought to be" some increasing function of the system's structural complexity. It would be nice if there were some general rule to the effect that, in any given complex system, the steady state which produces entropy at the maximum rate would at the same time be the steady state of maximum order and minimum entropy (Paltridge 2009).

Computer simulations by Damasceno, Engel and Glotzer (2012) show that the property entropy, a tendency generally described as "disorder" can nudge particles to form organized structures. By analyzing the shapes of the particles beforehand, they can even predict what kinds of structures will form (University of Michigan 2012).

Earlier studies on the application of the concept of maximum entropy in atmospheric physics are given below. A systems theory approach based on maximum entropy principle has been applied in cloud physics to obtain useful information on droplet size distributions without regard to the details of individual droplets (Liu et al. 1995; Liu 1995; Liu and Hallett 1997, 1998; Liu and Daum, 2001; Liu, Daum and Hallett 2002; Liu, Daum, Chai and Liu 2002). Liu, Daum et al. (2002) conclude that a combination of the systems idea with multiscale approaches seems to be a promising avenue. Checa and Tapiador (2011) have presented a maximum entropy approach to Rain Drop Size Distribution (RDSD) modelling. Liu, Liu and Wang (2011) have given a review of the concept of entropy and its relevant principles, on the organization of atmospheric systems and the principle of the Second Law of thermodynamics, as well as their applications to atmospheric sciences. The Maximum Entropy Production Principle (MEPP), at least as used in climate science, was first hypothesized by Paltridge (1978).

## 5. Cumulus Cloud Model

### 5.1 Introduction

Knowledge of the cloud dynamical, microphysical and electrical parameters and their interactions are essential for the understanding of the formation of rain in warm clouds and their modification. Extensive aircraft observations of cloud dynamical, microphysical and electrical parameters have been made in more than 2000 isolated warm cumulus clouds forming during the summer monsoon seasons (June-September) in Pune (18º 32'N, 73º 51'E, 559m asl), India (Selvam et al. 1980; Murty et al. 1975). The observations were made during aircraft traverses at about 300m above the cloud base. These observations have provided new evidence relating to the dynamics of monsoon clouds. A brief summary of the important results is given in the following. (i) Horizontal structure of the air flow inside the cloud has consistent variations with successive positive and negative values of vertical velocity representative of ascending and descending air currents inside the cloud. (ii) Regions of ascending currents are associated with higher liquid water content (LWC) and negative cloud drop charges and the regions of descending current are associated with lower LWC and positive cloud drop charges. (iii) Width of the ascending and descending currents is about 100m. The ascending and descending currents are hypothesized to be due to cloud-top-gravity oscillations. The cloud-top-gravity oscillations are generated by the intensification of turbulent eddies due to the buoyant production of energy by the microscale-fractional-condensation (MFC) in turbulent eddies. (iv) Measured LWC ($q$) at the cloud-base levels is smaller than the adiabatic value ($q_a$) with $q/q_a$ =0.6. The LWC increases with height from the base of the cloud and decreases towards the cloud-top regions. (v) Cloud electrical activity is found to increase with the cloud LWC. (vi) Cloud-drop spectra are unimodal near the cloud-base and multi-modal at higher levels. The variations in mean volume diameter (MVD) are similar to those in the LWC. (vii) In-cloud temperatures are colder than the environment. (viii) The lapse rates of the temperatures inside the cloud are less than the immediate environment. Environmental lapse rates are equal to the saturated adiabatic value. (ix) Increments in the LWC are associated with increments in the temperature inside the cloud. The increments in temperature are associated with the increments in temperature of the immediate environment at the same level or the level immediately above. (x) Variances of in-cloud temperature and humidity are higher in the regions where the values of LWC are higher. The variances of temperature and humidity are larger in the clear-air environment than in the cloud-air.

The dynamical and physical characteristics of monsoon clouds described above cannot be explained by simple entraining cloud models. A simple cumulus cloud model which can explain the observed cloud characteristics has been developed. The relevant physical concept and theory relating to dynamics of atmospheric planetary boundary layer (PBL), formation of warm cumulus clouds and their modification through hygroscopic particle seeding are presented in the following.

The mechanism of large eddy growth discussed in Section 2 in the atmospheric ABL can be applied to the formulation of the governing equations for cumulus cloud growth. Based on the above theory equations are derived for the in-cloud vertical profiles of (i) ratio of actual cloud liquid water content ($q$) to the adiabatic liquid water content ($q_a$) (ii) vertical velocity (iii) temperature excess (iv) temperature lapse rate (v) total liquid water content ($q_t$) (vi) cloud growth time (vii) cloud drop size spectrum (viii) rain drop size spectrum. The equations are derived starting from the microscale fractional condensation (MFC) process at cloud base levels. This provides the basic energy input for the total cloud growth.

## 5.2 Vertical profile of $q/q_a$

The observations of cloud liquid water content $q$ indicate that the ratio $q/q_a$ is less than one due to dilution by vertical mixing. The fractional volume dilution rate $f$ in the cloud updraft can be computed (Selvam 2013) from (8) (see Section 2.2.3) given by

$$f = \sqrt{\frac{2}{\pi z} \ln z}$$

In the above equation $f$ represents the fraction of the air mass of the surface origin which reaches the height $z$ after dilution by vertical mixing caused by the turbulent eddy fluctuations.

Considering that the cloud base level is 1000m the value of $R = 1000$m and the value of turbulence length scale $r$ below cloud base is equal to 100m so that the normalized length scale $z = R/r$ =1000m/100m=10 and the corresponding fractional volume dilution $f$ =0.6.

The value of $q/q_a$ at the cloud base level is also found to be about 0.6 by several observers (Warner 1970).

The fractional volume dilution $f$ will also represent the ratio $q/q_a$ inside the cloud. The observed (Warner 1970) $q/q_a$ profile inside the cloud is seen to closely follow the profile obtained by the model for dominant eddy radius $r$ =1m (Figure 3). It is therefore, inferred that, inside the cloud, the dominant turbulent eddy radius is 1m while below the cloud base the dominant turbulent eddy radius is 100m.

## *5.3 In-cloud vertical velocity profile*

The logarithmic wind profile relationship (4) derived for the PBL in Section 2.2.2 holds good for conditions inside a cloud because the same basic physical process, namely *Microscale Fractional Condensation* (MFC) operates in both the cases. The value of vertical velocity inside the cloud will however be much higher than in cloud-free air.

From (6) the in-cloud vertical velocity profile can be expressed as

$$W = w_* fz$$

Where

$W$ = vertical velocity at height $z$

$w_*$ = production of vertical velocity per second by the microscale-fractional-condensation at the reference level, i.e., cloud-base level

$f$ = fractional upward mass flux of air at level $z$ originating from the cloud-base level

The $f$ profile is shown in Figure 3. The vertical velocity profile will follow the $fz$ profile assuming $w_*$ is constant at the cloud-base level during the cloud growth period.

## 5.4 In-cloud excess temperature perturbation profile

The relationship between temperature perturbation θ and the corresponding vertical velocity perturbation is given as follows

$$W = \frac{g}{\theta_0}\theta$$

Where

g = acceleration due to gravity

$\theta_0$ = reference level potential temperature at the cloud-base level

By substituting for $W$ and taking $\theta_*$ as the production of temperature perturbation at the cloud-base level by microscale fractional condensation (MFC), we arrive at the following expression since there is a linear relationship between the vertical velocity perturbation $W$ and temperature perturbation θ (from 4 and 6)

$$\theta = \frac{\theta_*}{k}\ln z = \theta_* fz \qquad (41)$$

Thus the in-cloud vertical velocity and temperature perturbation follow the $fz$ distribution (Figure 4).

## 5.5 In-cloud temperature lapse rate profile

The saturated adiabatic lapse rate $\Gamma_{sat}$ is expressed as

$$\Gamma_{sat} = \Gamma - \frac{L}{C_p}\frac{d\chi}{dz}$$

Where

$\Gamma$ = dry adiabatic lapse rate

$C_p$ = specific heat of air at constant pressure

$d\chi/dz$ = liquid water condensed during parcel ascent along a saturated adiabat $\Gamma_{sat}$ in a height interval $dz$

In the case of cloud growth with vertical mixing the in-cloud lapse rate $\Gamma_s$ can be written as

$$\Gamma_s = \Gamma - \frac{L}{C_p}\frac{dq}{dz}$$

Where $dq$ which is less than $d\chi$ is the liquid water condensed during a parcel ascent $dz$. $q$ is less than the adiabatic liquid water content $q_a$. From (41)

$$\Gamma_s = \Gamma - \frac{d\theta}{dz} = \Gamma - \frac{\theta}{r} = \Gamma - \frac{\theta_* fz}{r} \qquad (42)$$

Where $d\theta$ is the temperature perturbation $\theta$ during parcel ascent $dz$. By concept $dz$ is the dominant turbulent eddy radius $r$ (Figure 2).

## 5.6 Total cloud liquid water content profile

The total cloud liquid water content $q_t$ at any level is directly proportional to $\theta$ as given by the following expression.

$$q_t = \frac{C_p}{L}\theta = \frac{C_p}{L}\theta_* fz = q_* fz \qquad (43)$$

Where $q_*$ is the production of liquid water content at the cloud-base level and is equal to $C_p\theta_*/L$. The total cloud liquid water content $q_t$ profile follows the $fz$ distribution (Figure 4).

*5.7 Cloud growth time*

The large eddy growth time (36) can be used to compute cloud growth time $T_c$

$$T_c = \frac{r_*}{w_*}\sqrt{\frac{\pi}{2}}\,\text{li}\left(\sqrt{z}\right)_{z_1}^{z_2} \qquad (44)$$

where li is the Soldner's integral or the logarithm integral. The cloud growth $T_c$ using (44) is shown in Figure 8.

## 6. Cloud model predictions and comparison with observations

Numerical computations of cloud parameters were performed for two different cloud base CCN mean volume radii, namely 2.2 μm and 2.5 μm and computed values are compared with the observations. The results are discussed below.

*6.1 Vertical velocity profile in the atmospheric boundary layer (ABL)*

The microscale fractional condensation (MFC) generated values of vertical velocity have been calculated for different heights above the surface for clear-air conditions and above the cloud-base for in-cloud conditions for a representative tropical environment with favourable moisture supply. A representative cloud-base height is considered to be 1000m above sea level (a.s.l) and the corresponding meteorological parameters are, surface pressure 1000 mb, surface temperature 30°C, relative humidity at the surface 80%, turbulent length scale 1 cm. The values of the latent heat of vapourisation $L_v$ and the specific heat of air at constant pressure $C_p$ are 600 cal $gm^{-1}$ and 0.24 cal $gm^{-1}$ respectively. The ratio values of $m_w/m_0$, where $m_0$ is the mass of the hygroscopic nuclei per unit volume of air and $m_w$ is the mass of water condensed on $m_0$, at various relative humidities as given by Winkler and Junge (1971, 1972) have been adopted and the value of $m_w/m_0$ is equal to about 3 for relative humidity 80%. For a representative value of $m_0$ equal to 100μg m$^{-3}$ the temperature perturbation $\theta'$ is equal to 0.00065°C and the corresponding vertical velocity perturbation (turbulent) $w_*$ is computed and is equal to 21.1x10$^{-4}$ cm sec$^{-1}$ from the following relationship between the corresponding virtual potential temperature $\theta_v$, and the acceleration due to gravity g equal to 980.6 cm sec$^{-2}$.

$$w_* = \frac{g}{\theta_v}\theta'$$

Heat generated by condensation of water equal to 300 μg on 100 μg of hygroscopic nuclei per meter$^3$, say in one second, generates vertical velocity perturbation $w_*$ (cm sec$^{-2}$) equal to 21.1x10$^{-4}$ cm sec$^{-2}$ at surface levels. Since the time duration for water vapour condensation by deliquescence is not known, in the following it is shown that a value of $w_*$ equal to 30x10$^{-7}$ cm sec$^{-2}$, i.e. about three orders of magnitude less than that shown in the above example is sufficient to generate clouds as observed in practice.

From the logarithmic wind profile relationship (4) and the steady state fractional upward mass flux $f$ of surface air at any height $z$ (8) the corresponding vertical velocity perturbation $W$ can be expressed in terms of the primary vertical velocity perturbation $w_*$ as (6)

$$W = w_* fz$$

$W$ may be expressed in terms of the scale ratio $z$ as given below

From (8) $$f = \sqrt{\frac{2}{\pi z}\ln z}$$

Therefore $$W = w_* z\sqrt{\frac{2}{\pi z}\ln z} = w_*\sqrt{\frac{2z}{\pi}\ln z}$$

The values of large eddy vertical velocity perturbation $W$ produced by the process of microscale fractional condensation at normalized height $z$ computed from (6) are given in the Table 2. The turbulence length scale $r_*$ is equal to 1 cm and the related vertical velocity perturbation $w_*$ is equal to $30 \times 10^{-7}$ cm/sec for the height interval 1cm to 1000m (cloud base level) for the computations shown in Table 2. Progressive growth of successively larger eddies generates a continuous spectrum of semi-permanent eddies anchored to the surface and with increasing circulation speed $W$.

Table 2. Vertical profile of eddy vertical velocity perturbation $W$

| Height above surface R | Length scale ratio z = R/r* | Vertical velocity W = w*fz cm sec-1 |
|---|---|---|
| 1 cm | 1 (r* = 1cm) | 30x10-7 (= w*) |
| 100 cm | 100 | 1.10x10-4 |
| 100 m | 100x100 | 2.20x10-3 |
| 1 km | 1000x100 | 8.71x10-3 ≈ 0.01 |
| 10 km | 10000x100 | 3.31x10-2 |

The above values of vertical velocity although small in magnitude are present for long enough time period in the lower levels and contribute for the formation and development of clouds as explained below.

### 6.2 Large eddy growth time

The time $T$ required for the large eddy of radius $R$ to grow from the primary turbulence scale radius $r_*$ is computed from (36) as follows.

$$T = \frac{r_*}{w_*} \sqrt{\frac{\pi}{2}} \int_{x1}^{x2} \mathrm{li}\left(\sqrt{z}\right)$$
$$x_1 = \sqrt{z_1} \text{ and } x_2 = \sqrt{z_2}$$

In the above equation $z_1$ and $z_2$ refer respectively to lower and upper limits of integration and li is the Soldner's integral or the logarithm integral. The large eddy growth time $T$ can be computed from (36) as follows.

As explained earlier, a continuous spectrum of eddies with progressively increasing speed (Table 2) anchored to the surface grow by microscale fractional condensation originating in turbulent fluctuations at the planetary surface. The eddy of radius 1000m has a circulation speed equal to 0.01 cm/sec (Table 2). The time $T$ seconds taken for the evolution of the 1000m ($10^5$ cm) eddy from 1cm height at the surface can be computed from the above equation by substituting for $z_1$ = 1cm and $z_2$ = $10^5$ cm such that $x_1=\sqrt{1}=1$ and $x_2=\sqrt{10^5}\approx317$.

$$T = \frac{1}{0.01}\sqrt{\frac{\pi}{2}}\int_{1}^{317} li(z)$$

The value of $\int_{1}^{317} li(z)$ is equal to 71.3

Hence $T \approx 8938$ sec $\approx$ 2 hrs 30 mins

Thus starting from the surface level cloud growth begins after a time period of 2 hrs 30 mins. This is consistent with the observations that under favourable synoptic conditions solar surface heating during the afternoon hours gives rise to cloud formation.

The dominant turbulent eddy radius at 1000m in the sub-cloud layer is 100 m starting from 1 cm radius dominant turbulent eddy at surface and formation of successively larger dominant eddies at decadic length scale intervals as explained in Section 2.2.1 above. Also, it has been shown in Section 5.2 that the radius of the dominant turbulent eddy ($r_*$) inside the cloud is 1m. These features suggest that the scale ratio is 100 times larger inside the cloud than below the cloud. The 1000m (1km) eddy at cloud base level forms the internal circulation for the next stage of eddy growth, namely 10km eddy radius with circulation speed equal to 0.03 cm/sec. Cloud growth begins at 1 km above the surface and inside this 10km eddy, with dominant turbulent eddy radius 1m as shown above. The circulation speed of this 1m radius eddy inside cloud is equal to 3m/sec as shown in the following. Since the eddy continuum ranging from 1cm to 10 km radius grows from the surface starting from the same primary eddy of radius $r_*$ and perturbation speed $w_*$ cm/sec, the circulation speeds of any two eddies of radii $R_1$, $R_2$ with corresponding circulation speeds $W_1$ and $W_2$ are related to each other as follows from (1).

$$W_1^2 = \sqrt{\frac{2}{\pi} \frac{r_*}{R_1}} w_*^2$$

$$W_2^2 = \sqrt{\frac{2}{\pi} \frac{r_*}{R_2}} w_*^2$$

$$\frac{W_2^2}{W_1^2} = \frac{R_1}{R_2}$$

$$\frac{W_2}{W_1} = \sqrt{\frac{R_1}{R_2}}$$

As mentioned earlier cloud growth with dominant turbulent eddy radius 1m begins at 1km above surface and forms the internal circulation to the 10km eddy. The circulation speed of the in-cloud dominant turbulent eddy is computed as equal to 3m/sec from the above equation where the subscripts 1 and 2 refer respectively to the outer 10km eddy and the internal 1m eddy.

The value of vertical velocity perturbation $W$ at cloud base is then equal to 100 times the vertical velocity perturbation just below the cloud base. Vertical velocity perturbation just below the cloud base is equal to 0.03 cm/sec from Table 2. Therefore the vertical velocity perturbation at cloud base is equal to 0.03x100cm/sec, i.e. 3cm/sec and is consistent with airborne observations over the Indian region during the monsoon season (Selvam et al. 1976; Pandithurai et al. 2011).

Cloud base vertical velocity equal to 1cm/sec has been used for the model computations in the following. The in-cloud updraft speed and cloud particle terminal velocities are given in Figure 9 for the two input cloud-base CCN size spectra with mean volume radius (mvr) equal to (i) 2.2 µm and (ii) 2.5 µm. The in-cloud updraft speed $W$ is the same for both CCN spectra since $W = w_* fz$ (6) and depends only on the persistent cloud base primary perturbation speed $w_*$ originating from micro-scale fractional condensation by deliquescence on hygroscopic nuclei at surface levels in humid environment (see Section 2). Cloud liquid water content increases with height (Figure 10) associated with increase in cloud particle mean volume radius (Figure 11) and terminal velocities (Figure 9). The cloud particles originating from larger size CCN (mvr=2.5 µm) are associated with larger cloud liquid water contents, larger mean volume radii and therefore larger terminal fall speeds at all levels.

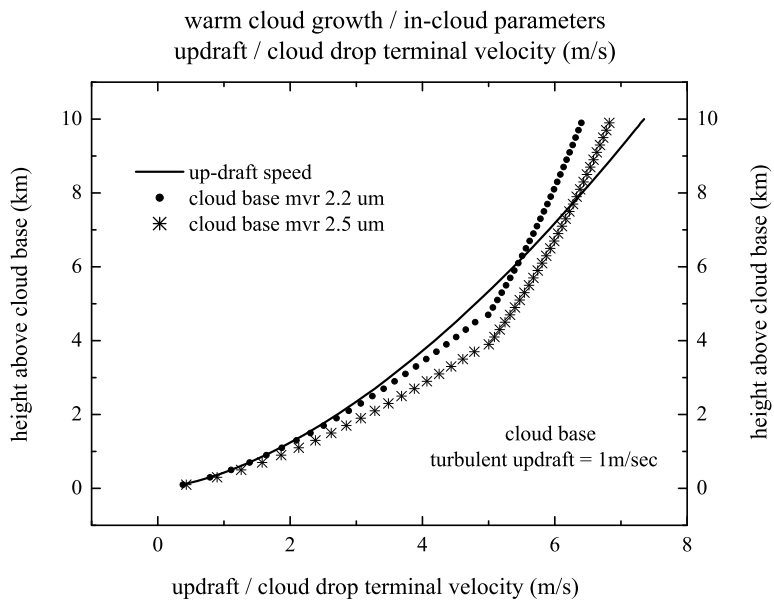

Figure 9. In-cloud updraft speed and cloud particle terminal velocities for the two input cloud-base CCN size spectra with mean volume radius (mvr) equal to (i) 2.2 µm and (ii) 2.5 µm.

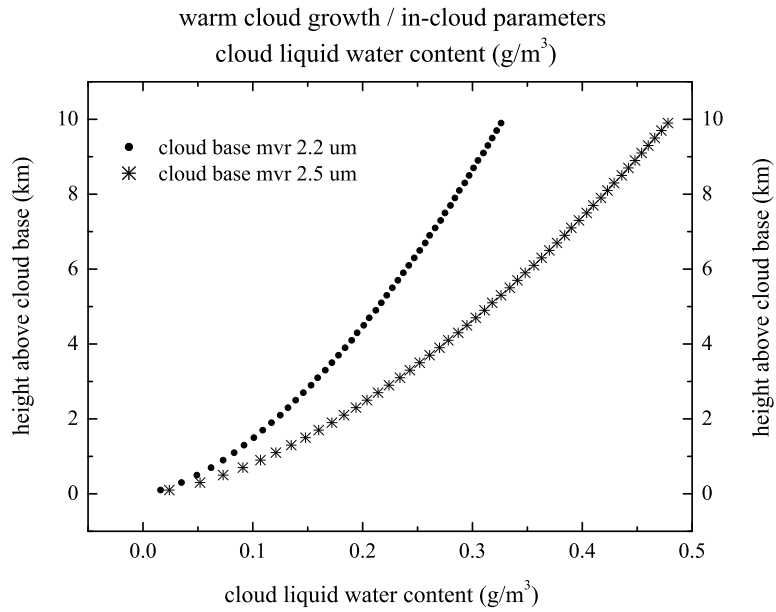

Figure 10. Cloud liquid water content

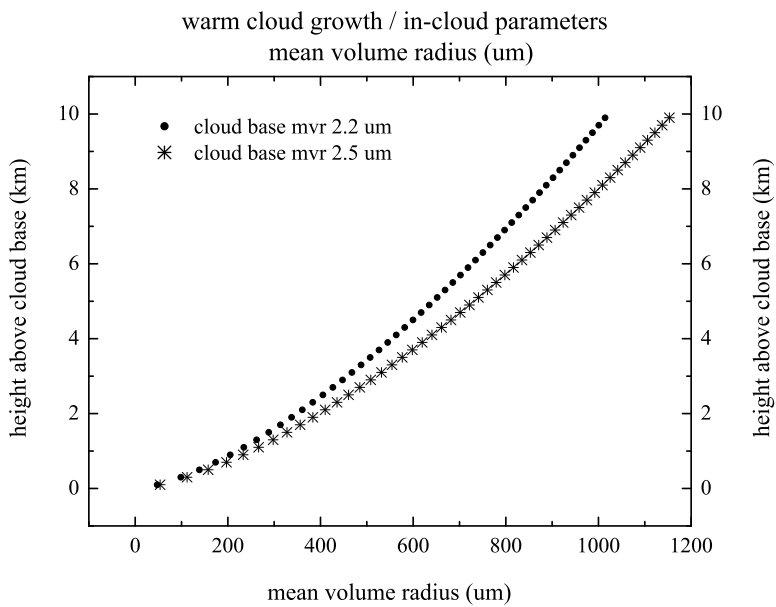

Figure 11: Cloud particle mean volume radius (μm)

The turbulent vertical velocity perturbation $w_*$ at cloud base level (1 km) is equal to .01m/sec or 1cm/sec. The corresponding cloud base temperature perturbation $\theta_*$ is then computed from the equation

$$w_* = \frac{g}{\theta_v}\theta_*$$

$$\theta_* = w_*\frac{\theta_v}{g}$$

*6.2.1 In-cloud temperature lapse rate*

The in-cloud lapse rate $\Gamma_s$ is computed using the following expression (42)

$$\Gamma_s = \Gamma - \frac{\theta_* fz}{r_*}$$

The primary eddy radius length $r_*$ at cloud base is equal to 1m as shown earlier (Section 5.2). The model computations for in-cloud vertical profile of vertical velocity $W$, temperature perturbation $\theta$ and lapse rate $\Gamma_c$ at 1 km height intervals above the cloud base and are given in Table 3.

The predicted temperature lapse rate decreases with height and becomes less than saturated adiabatic lapse rate near the cloud-top, the in-cloud temperatures being warmer than the environment. These results are in agreement with the observations.

Table 3. In-cloud vertical velocity and temperature lapse rates

| s.no | Height above cloud base $R$ (m) | Scale ratio $z = R/r_*$, ($r_*=1$ m) | $f$ | $fz$ | In-cloud vertical velocity (ms$^{-1}$) $W = w_*fz$ ms$^{-1}$, $w_* = .01$ ms$^{-1}$ | In-cloud temperature perturbation C $\theta = (\theta_* fz)$ C $\theta_*=0.00309$C | In-cloud lapse rate $\Gamma_c$ C/km $\Gamma_c = \Gamma-\theta$ $\Gamma = -10$C/km |
|---|---|---|---|---|---|---|---|
| 1 | 1000 | 1000 | .17426 | 174.26 | 1.74 | .538 | -9.46 |
| 2 | 2000 | 2000 | .13558 | 271.16 | 2.71 | .838 | -9.16 |
| 3 | 3000 | 3000 | .11661 | 349.82 | 3.50 | 1.081 | -8.92 |
| 4 | 4000 | 4000 | .10461 | 418.46 | 4.18 | 1.293 | -8.70 |
| 5 | 5000 | 5000 | .09609 | 480.43 | 4.80 | 1.484 | -8.52 |
| 6 | 6000 | 6000 | .08959 | 537.56 | 5.38 | 1.661 | -8.34 |
| 7 | 7000 | 7000 | .08442 | 590.91 | 5.91 | 1.826 | -8.17 |
| 8 | 8000 | 8000 | .08016 | 641.24 | 6.41 | 1.981 | -8.01 |
| 9 | 9000 | 9000 | .07656 | 689.05 | 6.89 | 2.129 | -7.87 |
| 10 | 10000 | 10000 | .07347 | 734.73 | 7.34 | 2.270 | -7.72 |

*6.2.2 Cloud growth time*

The large eddy growth time (36) can be used to compute cloud growth time $T_c$ (44)

$$T_c = \frac{r_*}{w_*}\sqrt{\frac{\pi}{2}}\,\text{li}\left(\sqrt{z}\right)\Big|_{z_1}^{z_2}$$

where li is the Soldner's integral or the logarithm integral. The vertical profile of cloud growth time $T_c$ is a function of only the cloud-base primary turbulent eddy fluctuations of radius $r_*$ and perturbation speed $w_*$ alone. The cloud growth $T_c$ using (44) is shown in Figure 12 for the two the different cloud-base CCN spectra, with mean volume radii equal to 2.2 μm and 2.5 μm respectively. The cloud growth time remains the same since the primary trigger for cloud growth is the persistent turbulent energy generation by condensation at the cloud-base in primary turbulent eddy fluctuations of radius $r_*$ and perturbation speed $w_*$.

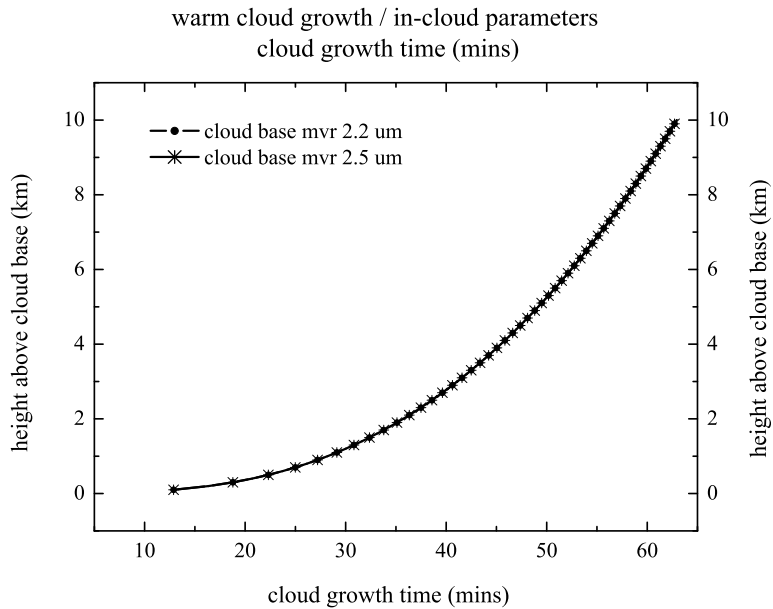

Figure 12: Cloud growth time

Considering $r_*$ is 100cm and $w_*$ is equal to 1cmsec$^{-1}$ (see Section 6.2) the time taken for cloud to grow to a height of e.g. 1600m above cloud base can be computed as shown below. The normalized height $z$ is equal to 1600 since dominant turbulent eddy radius is equal to 1m.

$$T_c = \frac{100}{100 \times 0.01}\sqrt{\frac{\pi}{2}} \text{li}\sqrt{1600}$$
$$= 100 \times 1.2536 \times 15.84 \text{ seconds}$$
$$\approx 30 \text{ minutes}$$

The above value is consistent with cloud growth time observed in practice.

*6.2.3 Cloud drop size spectrum*

The evolution of cloud drop size spectrum is critically dependent on the water vapour available for condensation and the nuclei number concentration in the sub-cloud layer. Cloud drops form as water vapour condenses in the air parcel ascending from cloud-base levels. Vertical mixing during ascent reduces the volume of cloud-base air reaching higher levels to a fraction *f* of its initial volume. Thus the nuclei available for condensation, i.e., the cloud drops number concentration also decreases with height according to the *f* distribution. The total cloud water content was earlier shown (Eq. 6) to increase with height according to the *fz* distribution. Thus bigger size cloud drops are formed on the lesser number of condensation nuclei available at higher levels in the cloud. Due to vertical mixing unsaturated conditions exist inside cloud. Water vapour condenses on fresh nuclei available at each level since, in the unsaturated in-cloud conditions micro-scale-fractional-condensation occurs preferentially on small condensation nuclei (Pruppacher and Klett 1997).

Earlier in Section 3 it was shown that the atmospheric eddy continuum fluctuations hold in suspension atmospheric particulates, namely, aerosols, cloud droplets and raindrops. The cloud drop size distribution spectrum also follows the universal spectrum derived earlier for atmospheric aerosol size distribution.

*6.2.4 In-cloud rain drop spectra*

In the cloud model it is assumed that bulk conversion of cloud water to rain water takes place mainly by collection and the process is efficient due to the rapid increase in the cloud water flux with height. The in-cloud raindrop size distribution also follows the universal spectrum derived earlier for suspended particulates in the atmosphere (Section 3).

The total rain water content $Q_r$ (c.c) is given as

$$Q_r = \frac{4}{3}\pi r_a^3 N = \frac{4}{3}\pi r_{as}^3 N f z^2$$

The above concept of raindrop formation is not dependent on the individual drop collision coalescence process.

Due to the rapid increase of cloud water flux with height, bulk conversion to rain water takes place in time intervals much shorter than the time required for the conventional collision-coalescence process.

The cloud base CCN size spectrum and the in-cloud particle (cloud and raindrop) size spectrum follow the universal spectrum (Figure 7) for suspended particulates in turbulent fluid flows. The cloud base CCN spectra and the in-cloud particulates (cloud and rain drops) size spectra at two levels 100m and 2km plotted in conventional manner as $dN/Nd(\log R)$ versus $R$ on log-log scale are shown in Figure 13. The in-cloud particulate size spectrum shifts rapidly towards larger sizes associated with rain formation. According to the general systems theory for suspended particulate size spectrum in turbulent fluid flows (Section 3), larger suspended particulates are associated with the turbulent regions (smaller scale length with larger fluctuation speed) of the vertical velocity spectrum. Spontaneous formation of larger cloud/rain drops may occur by collision and coalescence of smaller drops in these regions of enhanced turbulence.

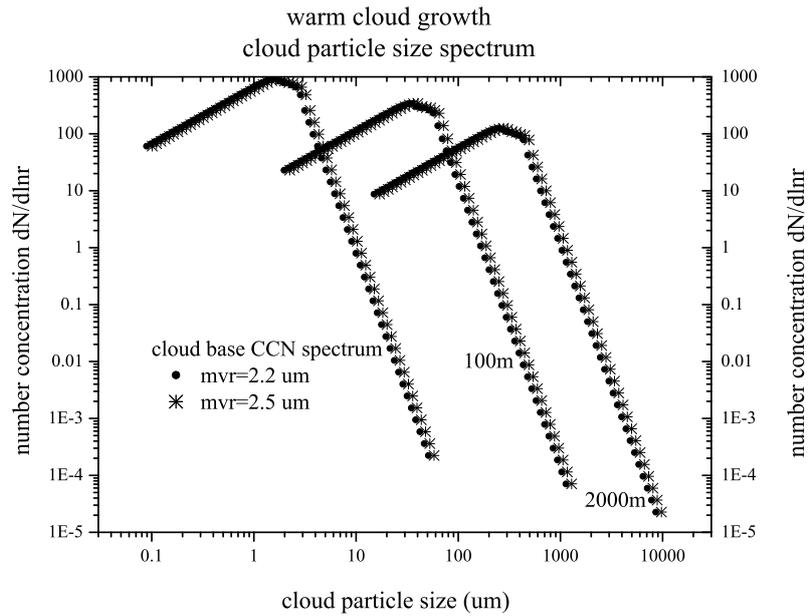

Figure 13: Cloud base CCN spectra and the in-cloud particulates (cloud and rain drops) size spectra at two levels 100m and 2km

## 7. Rainfall commencement

Rainfall sets in at the height at which the terminal velocity $w_T$ of the rain drop becomes equal to the mean cloud updraft $W$. Let the mean volume radius $R_m$ be representative of the precipitation drop at level $z$ above cloud base. In the intermediate range (40 μm $< R_m <$ 0.6 mm), an approximated formula for the fall speed is given by Yau and Rogers (1989) as $w_T = K_3 R_m$ where $K_3 = 8 \times 10^3$ sec$^{-1}$.

For droplets with radii < 30 μm, the terminal velocity is given as $w_T = K_1 R_m^2$ where $K_1 \sim 1.19 \times 10^6$ cm$^{-1}$ s$^{-1}$.

For large drops, $w_T = K_2 \sqrt{R_m}$. $K_2$ can be approximated as $K_2 = 2010 \text{cm}^{1/2} \text{ s}^{-1}$ for a raindrop size between 0.6mm and 2 mm (Yau and Rogers 1989).

*7.1 Rainfall rate*

The in-cloud rainfall rate ($R_t$) can be computed as shown below.

Rain water production rate over unit area (rainfall rate mm/sec/unit area) $Q_{rt}$ in the cloud at any level $z$ above the cloud-base (45) is given by

$$R_{tz} = (W_T - W)\frac{4}{3}\pi R_m^3 N$$

The in-cloud rainfall rate $R_{tz}$ is derived from the normalised flux value for height $z$, i.e. eddy wavelength $z$. Therefore for unit normalized height interval $z$, the rainfall rate $R_t$ is equal to $R_{tz}/z$ cm/sec. $R_t = (R_{tz}/z)*3600.*10$ mm/hr.

In the above equation $W_T$ in cm/sec is the terminal velocity of the raindrop of mean volume radius $R_m$ at level $z$.

Rainfall rate ($R_t$) derived above is representative of the mean cloud scale rain intensity.

Rain formation computed for two different cloud base CCN mean volume radii, namely 2.2 µm and 2.5 µm are shown in Figure 14. Rain with larger rain rate forms at a lower level and extends up to a higher level for the larger size (mvr 2.5 µm) cloud-base CCN spectrum.

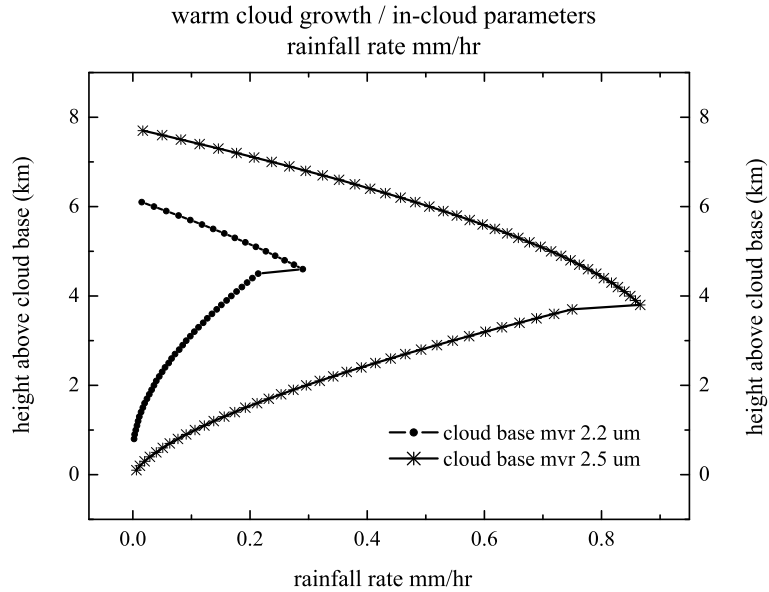

Figure 14: In-cloud rainrate (mm/hr)

**8. Warm Cloud Responses to Hygroscopic Particle Seeding**

In Sections 2, and 3 it is shown that the buoyant production of energy released by condensation in turbulent eddies is mainly responsible for the formation and growth of the cloud. When warm clouds are seeded with hygroscopic particles the turbulent buoyant energy production increases due to condensation and results in enhancement of vertical mass exchange. This would result in enhancement of convergence in the sub-cloud layer and invigoration of the updraft in the cloud. If sufficient moisture is available in the sub-cloud air layer the enhanced convergence would lead to increased condensation and cloud growth. According to the above physical concept and theory of the cumulus cloud model presented in this paper it can be concluded that hygroscopic particle seeding alters the dynamics of warm clouds.

*8.1 Dynamic effect of salt seeding*

Woodcock and Spencer (1967) hypothesized that dispersion of NaCl particles in a nearly water saturated atmosphere would be sufficient to initiate a cumulus cloud. Their experiments have produced visible clouds when dry NaCl particles in the size range 0.5 to 20 μm diameter, were released from aircraft in the warm moist marine boundary layer near Hawaii. The temperatures in the salt laden air were about 0.4°C higher than that of the environment on the average. Observations of the possible dynamic effects in warm clouds by hygroscopic particle seeding have been recorded (Murty et al. 1975). The dynamic effect of salt seeding has been estimated from the following calculations which are based on the theory and the physical concept of the cumulus cloud model discussed earlier.

The latent heat released by condensation on the hygroscopic nuclei is computed for the turbulent vertical velocities of 1m/sec and the turbulent eddy radius of 50m. The relative humidity inside the cloud at the seeding level near the cloud base is assumed to be equal to 90 per cent. A seeding rate of 500 gm/sec and a volume dispersion rate of 45000 m$^3$/sec are assumed. The value of the seeding rate assumed is based on that used in the aircraft warm cloud modification experiment in India (Murty et al. 1975). The value of the dispersion rate used in the computation is equal to half the value assumed by Woodcock and Spencer (1967). The median volume diameter of the salt nuclei is 10 μm. One such nucleus gives rise to the formation of a water droplet of radius 9.5 μm in a time period of 1 second at the assumed relative humidity of 90% inside the cloud (Keith and Aarons 1954). Thus the mass of water condensed on the seed nuclei per second per cubic meter of seeded volume can be computed. The latent heat released by this condensation will give rise to a heating rate of 0.09°Cm$^{-3}$sec$^{-1}$ of the seeded volume. The temperature increase will result in a vertical velocity increase of about 0.3cmsec$^{-2}$. This increase in the vertical velocity occurs in successive 50m lengths of the cloud along the flight path of the aircraft whose speed is 50msec$^{-1}$. The mean increase in the turbulent vertical velocity along a 500m length of the cloud can be approximately taken to be equal to one tenth of the vertical velocity in individual 50m length of aircraft path, i.e., equal to 0.03cmsec$^{-2}$. Therefore, the mean increase in the vertical velocity for the total path length of 500m, i.e., the large eddy updraft region is equal to one fourth of the mean increase in turbulent vertical velocity (1). Thus the vertical velocity perturbation in large eddy updraft region is equal to about 0.01 cmsec$^{-2}$. This cloud scale vertical velocity perturbation due to the hygroscopic particle seeding is equal to the naturally occurring vertical velocity perturbation at the cloud-base level for the sample cloud case discussed in Sections 6 and 7. Also, it was shown that the cloud–base vertical velocity production is the main driving agent for the cloud growth processes. The sample computations of the dynamic effect of salt seeding discussed above indicated that the total cloud growth processes in the case of seeded clouds would be increased by 100%. Evidence for such dynamic effects due to salt seeding in warm cumulus clouds can be seen from the observation of cloud liquid water content, in-cloud temperature and cloud-top growth rates made in seeded clouds (Murty et al. 1975).

## 9. Conclusions

Atmospheric flows exhibit selfsimilar fractal fluctuations, a signature of long-range correlations on all space-time scales. Realistic simulation and prediction of atmospheric flows requires the incorporation of the physics of observed fractal fluctuation characteristics in traditional meteorological theory. A general systems theory model for fractal space-time fluctuations in turbulent atmospheric flows (Selvam 2013) is presented. Classical statistical physical concepts underlie the physical hypothesis relating to the dynamics of the atmospheric eddy systems and satisfy the maximum entropy principle of statistical physics. The model predictions are as follows.

i. Fractal fluctuations of atmospheric flows signify an underlying eddy continuum with overall logarithmic spiral trajectory enclosing a nested continuum of vortex roll circulations which trace the quasiperiodic Penrose tiling pattern. Satellite images of cyclones and hurricanes show the vivid logarithmic spiral cloud pattern whose precise mathematical equiangular spiral (golden) geometry has been used by meteorologists (Senn et al. 1957; Senn and Hisar 1959; Sivaramakrishnan and Selvam 1966; Wong, Yip and Li 2008) for locating the center (eye) of the storm.

ii. The logarithmic law of wall for boundary layer flows is derived as a natural consequence of eddy mixing process. The von karmen's constant is equal to $1/\tau^2$ ($\approx 0.38$) where $\tau$ is the golden mean ($\approx 0.618$).

iii. The probability distribution for amplitude as well as variance of fractal fluctuations of meteorological parameters are given by the same universal inverse power law function $P$, namely $P = \tau^{-4t}$ where the normalized standard deviation $t$ designates the eddy length step growth number. Such a result that the additive amplitudes of eddies when squared represent the probability densities of the fluctuations is observed in the subatomic dynamics of quantum systems. Therefore fractal fluctuations are signatures of quantum-like chaos in dynamical systems.

iv. The probability distribution $P$ of amplitudes of fractal fluctuations is close to the statistical normal distribution for values of normalized standard deviation $t$ values equal to or less than 2. The probability of occurrence of extreme events, i. e., normalized deviation $t$ values greater than 2, is close to zero as given by the statistical normal distribution while $P$ distribution for fractal fluctuations gives appreciably high values as observed in practice.

v. Atmospheric eddy energy (variance) spectrum follows the universal inverse power law form $P = \tau^{-4t}$ indicating long-range space-time correlations between local (small-scale) and global (large scale) perturbations.

vi. Atmospheric eddy energy (variance) spectrum follows the universal inverse power law form $P = \tau^{-4t}$ indicating long-range space-time correlations between local (small-scale) and global (large scale) perturbations.

vii. Numerical computations of cloud parameters were performed for two different cloud base CCN mean volume radii, namely 2.2 μm and 2.5 μm and computed values are compared with the observations. Cloud–base vertical velocity production by *microscale fractional condensation* is the main driving agent for the cloud growth processes. The cloud growth time is about 30 mins as observed in practice (McGraw and Liu 2003) and is the same for the two CCN spectra since the primary trigger for cloud growth is the persistent turbulent energy generation by condensation at the cloud-base in primary turbulent eddy fluctuations of radius $r_*$ and perturbation speed $w_*$. However, for the larger CCN mean volume radius, namely 2.5 μm, raindrops form earlier at a lower level and extend upto higher levels in the cloud. Under suitable conditions of humidity and moisture in the environment warm rain formation can occur at as short a time interval as 30 mins.

viii. Hygroscopic particle seeding alters the dynamics of warm clouds and enhances rainfall upto 100% under favourable conditions of moisture supply in the environment.

**Acknowledgement**

The author is grateful to Dr.A.S.R.Murty for encouragement during this study.

# References

Bak, P.C., Tang, C., and Wiesenfeld, K. (1988), 'Self-organized criticality', *Phys. Rev. A.*, 38, 364-374.

Beck, C. (2009), 'Generalized information and entropy measures in physics', *Contemp. Phys.*, 50, 495-510. arXiv:0902.1235v2 [cond-mat.stat-mech]

Boers, R. (1989), 'A parametrization of the depth of the entrainment zone', *J. Atmos. Sci.*, 28, 107-111.

Brown, R.A. (1980), 'Longitudinal instabilities and secondary flows in the planetary boundary layer', *Rev. Geophys. Space Phys.*, 18, 683-697.

Chakrabarti, C.G., and De, K. (2000), 'Boltzmann-Gibbs entropy: axiomatic characterization and application', *Internat. J. Math. & Math. Sci.*, 23(4), 243-251.

Checa, R., and Tapiador, F.J. (2011), 'A maximum entropy modelling of the rain drop size distribution', *Entropy* 13, 293-315.

Damasceno, P., Engel, M., and Glotzer, S. (2012) 'Predictive self-assembly of polyhedra into complex structures', *Science*, 337(6093), 453-457.

Donnelly, R.J. (1988) 'Superfluid turbulence', *Sci. Am.*, 259 (5), 100-109.

Donnelly, R.J. (1990), *Quantized Vortices in Helium II*. USA: Cambridge University Press.

Eady, E.T. (1950), 'The cause of the general circulation of the atmosphere', *Cent. Proc. Roy. Met. Soc.* 1950, aos.princeton.edu.

Grabowskii, W.W., and Wang, L-P. (2012), 'Growth of cloud droplets in a turbulent environment', *Annu. Rev. Fluid Mech.* DOI: 10.1146/annurev-fluid-011212-140750

Gryning, S-E., and Batchvarova, E. (2006), 'Parametrization of the depth of the entrainment zone above the daytime mixed layer', *Quart. J. Roy. Meteor. Soc.*, 120 (515), 47-58.

Haken, H. (1977), *Synergetics, an Introduction: Nonequilibrium Phase Transitions and Self-Organization in Physics, Chemistry, and Biology*, New York: Springer-Verlag.

Holton, J.R. (2004), *An Introduction to Dynamic Meteorology*, USA: Academic Press.

Hultmark, M., Vallikivi, M., Bailey, S., and Smits, A. (2012), 'Turbulent pipe flow at extreme Reynolds numbers', *Phys. Rev. Lett.*, 108(9), 094501.

Kaniadakis, G. (2009), 'Maximum entropy principle and power-law tailed distributions', *Eur. Phys. J. B*, 70, 3-13.

Keith, C.H., and Arons, A.B. (1954), 'The growth of sea-salt particles by condensation of atmospheric water vapour', *J. Meteor.*, 11, 173-184.

Klir, G.J. (2001, 1991), *Facets of Systems Science*, Second Edition. IFSR International Series on Systems Science and Engineering, Vol 15, NewYork: Kluwer Academic/Plenum Publishers.

Lawrence Berkeley National Laboratory (2013), How computers push on the molecules they simulate. http://phys.org/news/2013-01-molecules-simulate.html. Accessed on 6 January 2013

Liu, V. (1956), 'Turbulent dispersion of dynamic particles', *J. Meteor.*, 13, 399-405.

Liu, Y. (1992), 'Skewness and Kurtosis of measured raindrop size distributions', *Atmos. Environ.*, 26A, 2713-2716.

Liu, Y. (1995) 'On the generalized theory of atmospheric particle systems', *Adv. Atmos. Sci.*, 12, 419-438.

Liu, Y., Laiguang, Y., Weinong, Y., and Feng, L. (1995), 'On the size distribution of cloud droplets', *Atmos. Res.*, 35, 201-216.

Liu, Y., and Hallett, J. (1997), 'The "1/3" power-law between effective radius and liquid water content', *Quart. J. Roy. Meteor. Soc.*, 123, 1789-1795.

Liu, Y., and Hallett, J. (1998) 'On size distributions of droplets growing by condensation: A new conceptual model', *J. Atmos. Sci.*, 55, 527-536.

Liu, Y., and Daum, P.H. (2001) 'Statistical physics, information theory and cloud droplet size distributions', *Eleventh ARM Science Team Meeting Proceedings*, Atlanta, Georgia, March 19-23.

Liu, Y., Daum, P.H., and Hallett, J. (2002), 'A generalized systems theory for the effect of varying fluctuations on cloud droplet size distributions', *J. Atmos. Sci.*, 59, 2279-2290.

Liu, Y., Daum, P.H., Chai, S.K., and Liu, F. (2002), 'Cloud parameterizations, cloud physics, and their connections: An overview', *Recent Res. Devel. Geophys.*, 4, 119-142.

Liu, Y., Liu, C., and Wang, D. (2011), 'Understanding atmospheric behaviour in terms of entropy: a review of applications of the second law of thermodynamics to meteorology', *Entropy*, 13, 211-240.

Lovejoy, S., and Schertzer, D. (2010), 'Towards a new synthesis for atmospheric dynamics: space-time cascades', *Atmos. Res.*, 96, 1-52. http://www.physics.mcgill.ca/~gang/eprints/eprintLovejoy/neweprint/Atmos.Res.8.3.10.finalsdarticle.pdf

Marusic, I., McKeon, B.J., Monkewitz, P.A., Nagib, H.M., Smits, A.J., and Sreenivasan, K.R. (2010), 'Wall-bounded turbulent flows at high Reynolds numbers: Recent advances and key issues', *Phys. Fluids*, 22, 065103.

McGraw, R., and Liu, Y. (2003), 'Kinetic potential and barrier crossing: A model for warm cloud drizzle formation', *Phys. Rev. Lett.*, 90(1), 018501-1.

Murty, A.S.R., Selvam, A.M., and Ramanamurty, Bh.V. (1975), 'Summary of observations indicating dynamic effect of salt seeding in warm cumulus clouds', *J. Appl. Meteor.*, 14, 629-637.

Newman, M.E.J. (2011), 'Complex Systems: A Survey', arXiv:1112.1440v1 [cond-mat.stat-mech] 6 Dec 2011

Paltridge, G.W. (1978), 'Climate and thermodynamic systems of maximum dissipation', *Nature*, 279, 630–631.

Paltridge, G.W. (2009), 'A story and a recommendation about the principle of maximum entropy production', *Entropy*, 11, 945-948.

Pandithurai, G., Dipu, S., Prabha, T.V., Maheskumar, R.S., Kulkarni, J.R., and Goswami, B.N. (2012), 'Aerosol effect on droplet spectral dispersion in warm continental cumuli', *J. Geophys. Res.*, 117, D16202.

Parmeggiani, A. (2012) 'Viewpoint', *Physics*, 5, 118 ('A Viewpoint on: Exact current statistics of the asymmetric simple exclusion process with open boundaries', Gorissen, M., Lazarescu, A., Mallick, K., and Vanderzande, C. (2012), *Phys. Rev. Lett.*, 109, 170601– *Physics*, 5, 118). DOI: 10.1103/Physics.5.118.

Peters, O., Hertlein, C., and Christensen, K. (2002), 'A complexity view of rainfall', *Phys. Rev. Lett.*, 88, 018701.

Peters, O., Deluca, A., Corral, A., Neelin, J.D., and Holloway, C.E. (2010), 'Universality of rain event size distributions', *J. Stat. Mech*. P11030.

Prandtl, L. (1932), 'Ergeb. Aerodyn. Versuch', *Gottingen*, 4, 18-29.

Pruppacher, H.R., and Klett J.D. (1997), *Microphysics of Clouds and Precipitation*, The Netherlands: Kluwer Academic Publishers.

Salingaros, N.A., and West, B.J. (1999), 'A universal rule for the distribution of sizes', *Environment and Planning B: Planning and Design*, 26, 909-923, Pion Publications.

Selvam, A.M., Murty, A.S.R., Vijayakumar, R., Paul, S.K., Manohar, G.K., Reddy, R.S., Mukherjee, B.K., and Ramanamurty, Bh.V. (1980), 'Some thermodynamical and microphysical aspects of monsoon clouds', *Proc. Indian Acad. Sci.*, 89A, 215-230.

Selvam, A.M. (1993) 'Universal quantification for deterministic chaos in dynamical systems', *Applied Mathematical Modelling*, 17, 642-649. http://xxx.lanl.gov/html/physics/0008010

Selvam, A.M. (2013), 'Scale-free universal spectrum for atmospheric aerosol size distribution for Davos, Mauna Loa and Izana', *International J. Bifurcation & Chaos*, 23(2), 1350028.

Senn, H.V., Hiser, H.W., and Bourret, R.C. (1957), 'Studies of hurricane spiral bands as observed on radar', Final Report, U. S. Weather Bureau Contract No. Cwb-9066, University of Miami, 21pp. (1957) [NTIS-PB-168367].

Senn, H.V., and Hiser, H.W. (1957), 'On the origin of hurricane spiral bands', *J. Meteor.*, 16, 419-426.

Sethna, J.P. (2009), *Statistical Mechanics: Entropy, Order Parameters, and Complexity*, Oxford: Clarendon Press. http://www.freebookcentre.net/physics-books-download/Statistical-Mechanics-Entropy,-Order-Parameters,-and-Complexity-[PDF-371].html

Sivak, D. A., Chodera, J. D., and Crooks, G. E. (2012), 'Using nonequilibrium fluctuation theorems to understand and correct errors in equilibrium and nonequilibrium discrete Langevin dynamics simulations', *Phys. Rev. X*, 3, 011007. arXiv:1107.2967v4 [cond-mat.stat-mech] 5 Oct 2012

Sivaramakrishnan, M.V., and Selvam, M.M. (1966), 'On the use of the spiral overlay technique for estimating the center positions of tropical cyclones from satellite photographs taken over the Indian region', *Proc. 12th conf. Radar Meteor.*, 440-446.

Skyttner, L. (2005), *General systems theory: Problems, Perspectives, Practice* (2nd Edition). Singapore: World Scientific Co. Pte. Ltd.

Townsend, A.A. (1956), *The Structure of Turbulent Shear Flow*, 2nd ed., London, U. K.: Cambridge University Press, pp.115-130.

Tuck, A.F. (2010), 'From molecules to meteorology via turbulent scale invariance', *Q. J. R. Meteor. Soc.*, 136, 1125–1144.

University of Michigan, (2012), 'Entropy can lead to order, paving the route to nanostructures', *ScienceDaily* Retrieved July 27, 2012, http://www.sciencedaily.com/releases/2012/07/120726142200.htm

Von Bertalanffy, L. (1972), 'The history and status of general systems theory', *The Academy of Management Journal*, 15(4), General Systems Theory, 407-426.

Von Karman, T. (1930), 'Nachr. Ges. Wiss. Gottingen', *Math. Phys. Kl.*, 58-76, 322.

Warner, J. (1970), 'The micro structure of cumulus clouds Part III: The nature of updraft', *J. Atmos. Sci.*, 27, 682-688.

Winkler, P., and Junge, C.E. (1971), 'Comments on "Anomalous deliquescence of sea spray aerosols"', *J. Appl. Meteor.*, 10, 159-163.

Winkler, P., and Junge, C. (1972), 'The growth of atmospheric aerosol particles as a function of the relative humidity', *J. de Recherches Atmospheriques*, 6, 617-637.

Wong, K.Y. Yip, C.L., and Li, P.W. (2008), 'Automatic identification of weather systems from numerical weather prediction data using genetic algorithm', *Expert Systems with Applications*, 35(1-2), 542-555.

Woodcock, A.H., and Spencer, A.T. (1967), 'Latent heat released experimentally by adding Sodium Chloride particles to the atmosphere', *J. Appl. Meteor.* 6, 95-101.

Yano, J-I. (2012) 'Self–organized criticality and homeostasis in atmospheric convective organization', *J. Atmos. Sci.,* 69, 3449-3462.

Yau, M.K., Rogers, R.R. (1989) *A Short Course in Cloud Physics*. Third Edition (Intl Ser. Nat. Phil.) U.S.A.: Butterworth-Heinemann.